\documentclass[12pt, twoside, openany, oldfontcommands]{memoir}
\nouppercaseheads
\setsecnumdepth{subsection}
\maxtocdepth{subsection}
\usepackage[a4paper,
			lmargin=1.5cm,
			rmargin=1.5cm,
			tmargin=1.5cm,
			bmargin=3.0cm,
			headsep=1cm,
			footskip=1cm,
			headheight=2mm,
			includefoot]{geometry}
\usepackage[dvipsnames]{xcolor}
\usepackage{graphicx}
\usepackage{float}
\usepackage{authblk}
\usepackage[T1]{fontenc}
\usepackage[utf8]{inputenc}
\usepackage{amsmath}
\usepackage{amssymb}
\usepackage{wasysym}
\usepackage{mathrsfs}
\usepackage{csquotes}
\usepackage{fancybox}
\usepackage[hidelinks]{hyperref}
\usepackage{braket}
\usepackage{soul}
\usepackage{tikz}
\usepackage{stmaryrd}
\usetikzlibrary{shapes,shadows}
\usepackage{cases}
\usepackage{subfig}
\usepackage{siunitx}
\usepackage{bbold}
\usepackage[most]{tcolorbox}
\usepackage[labelnumber,defernumbers,style=alphabetic,natbib=true,backend=biber,maxalphanames=1,minalphanames=1]{biblatex}
\def\flagdraft{false}

\makeoddfoot{ruled}{}{\thepage}{} 
\makeevenfoot{ruled}{}{\thepage}{} 

\colorlet{color_inc}  {black}
\colorlet{color_sub}  {RoyalBlue!100!}
\colorlet{color_cov}  {JungleGreen!100!}
\colorlet{color_embed}{GreenYellow!100!}
\colorlet{color_dep}  {Goldenrod!100!}
\colorlet{color_sup}  {ProcessBlue!100!}
\colorlet{color_rod}  {Peach!100!}
\colorlet{color_refl} {OrangeRed}
\colorlet{color_tran} {SkyBlue}

\newtcolorbox{boxgrey}[2][]{
    lower separated=false,
    colback=white!90!gray,
    colframe=white, fonttitle=\bfseries,
    colbacktitle=white!70!gray,
    coltitle=black,
    enhanced,
    attach boxed title to top left={xshift=0.5cm,yshift=-2mm},
    title={#2},#1
}

\newtcolorbox{boxgrey2}[2][]{
    lower separated=false,
    colback=white!90!gray,
    colframe=white,
    colbacktitle=white!70!gray,
    coltitle=black,
    enhanced,
    attach boxed title to top left={xshift=0.cm,yshift=-2mm},
    title={#2},#1
}

\newtcolorbox{boxblack}[2][]{
    lower separated=false,
    colback=white,
    colframe=black,fonttitle=\bfseries,
    colbacktitle=black,
    coltitle=white,
    enhanced,
    attach boxed title to top left={yshift=-0.1in,xshift=0.15in},
    boxed title style={boxrule=0pt,colframe=white,},
    title=#2,#1
}

\def\x{\mathbf{x}}
\def\n{\mathbf{n}}
\def\y{\mathbf{y}}
\def\z{\mathbf{z}}
\def\rr{\mathbf{r}}

\def\build#1_#2^#3{\mathrel{\mathop{\kern 0pt#1}\limits_{#2}^{#3}}}

\def\curle{\ensuremath{\mathbf{curl}}}

\def\curlet{\ensuremath{\mathbf{curl}_t}}
\def\div{\mathrm{div}\,}
\def\grad{\ensuremath{\mathbf{grad}}\,}
\def\gradt{\ensuremath{\mathbf{grad}_t}}

\def\bk{\ensuremath{\mathbf{k}}}
\def\bE{\ensuremath{\mathbf{E}}}

\def\bW{\ensuremath{\mathbf{W}}}

\def\bx{\ensuremath{\mathbf{r}}}
\def\bx{\ensuremath{\mathbf{x}}}

\def\bsh{\ensuremath{\hat{\mathbf{s}}}}
\def\bph{\ensuremath{\hat{\mathbf{p}}}}
\def\bxh{\ensuremath{\hat{\mathbf{x}}}}
\def\byh{\ensuremath{\hat{\mathbf{y}}}}
\def\bzh{\ensuremath{\hat{\mathbf{z}}}}

\def\exp{\ensuremath{\mathrm{exp}}}
\def\cos{\ensuremath{\mathrm{cos}}}
\def\sin{\ensuremath{\mathrm{sin}}}

\newcommand{\epsrin}[1]{\ensuremath\varepsilon_{r,#1}}
\newcommand{\murin}[1]{\ensuremath\mu_{r,#1}}

\def\bEinc{\mathrm{\mathbf{E}}^{\mathrm{inc}}}
\def\bHinc{\mathrm{\mathbf{H}}^{\mathrm{inc}}}
\newcommand{\bEd}[1]{\ensuremath{\mathbf{E}_{#1}^{\mathrm{d}}}}
\def\om{\omega}

\def\d{\ensuremath{\,\mathrm{d}}}

\def\bE{\ensuremath{\mathbf{E}}}
\def\br{\ensuremath{\boldsymbol{r}}}
\def\bH{\ensuremath{\mathbf{H}}}
\def\bk{\ensuremath{\mathbf{k}}}

\def\bF{\ensuremath{\boldsymbol{F}}}

\def\div{\mathop{\rm div}\nolimits}

\def\N{\mathbb N}

\def\grad{\mathop{\rm \mathbf{grad}}\nolimits}
\def\div{\mathop{\rm div}\nolimits}
\def\curl{\mathop{\rm  \mathbf{curl}}\nolimits}

\def\cos{\mathop{\rm cos}\nolimits}
\def\sin{\mathop{\rm sin}\nolimits}
\def\tan{\mathop{\rm tan}\nolimits}

\newcommand{\tensid}{\boldsymbol{\mathsf{I}}_\mathsf{d}}

\newcommand{\tensdelta}{\boldsymbol{\delta}}
\newcommand{\tensepst}{{\boldsymbol{\widetilde \varepsilon}}}
\newcommand{\tensmut}{{\boldsymbol{\widetilde \mu}}}
\newcommand{\tensdeltat}{{\boldsymbol{\widetilde \delta}}}

\newcommand{\tensxi}{\boldsymbol{\xi}}

\newcommand{\GalerkinV}[3]{\ensuremath{\displaystyle\int_{#3}#1\cdot\overline{#2}\,\mathrm{d}{\Omega}}}
\newcommand{\GalerkinS}[3]{\ensuremath{\displaystyle\int_{#3}#1\cdot\overline{#2}\,\mathrm{d}{\Gamma}}}

\def\tensmur{\boldsymbol{\mu}_r}

\def\tensepsr{\boldsymbol{\varepsilon}_r}
\def\tensepsra{\boldsymbol{\varepsilon}_{r,a}}
\def\tensmurtt{\boldsymbol{\tilde{\mu}}_{r}}
\def\tensepsrtt{\boldsymbol{\tilde{\varepsilon}}_{r}}
\def\tensepsratt{\boldsymbol{\tilde{\varepsilon}}_{r,a}}
\def\tensmura{\boldsymbol{\mu}_{r,a}}
\def\tensid{\mathbb{1}}
\def\dfrac{\displaystyle\frac}
\def\dint{\displaystyle\int}

\newcommand{\bashline}[1]{{
    \colorlet{foo}{gray!10}
    \sethlcolor{foo}
    \noindent{\texttt{\footnotesize{\hl{#1}}}}
  }
}

\newcommand{\RE}{\Re {\it{e}}}
\newcommand{\IM}{\Im {\it{m}}}

\newcommand*{\stt}[1]{{\small \tt #1}}
\newcommand*{\mybox}[1]{\framebox{\strut #1}}
\newcommand{\checkboxC} {\makebox[0pt][l]{$\square$}\raisebox{.15ex}{\hspace{0.1em}$\checkmark$}}

\newcommand{\button}[2]
	{\begin{tikzpicture}[baseline=(tempname.base)]
	      \node[draw=black!60,
							fill=#1, 
							rounded corners=0.5pt, 
							inner sep=0pt, 
							minimum width=0.5em, 
							minimum height=1em, 
							general shadow={fill=black, shadow xshift=0.1pt, shadow yshift=-0.1pt, shadow scale=1.1}] 
							(tempname) {#2};
	  \end{tikzpicture}}

\newcommand{\onelabbuttons}[2]
	{	\makebox[16.5pt][l]{\makebox[5.7pt][l]{\raisebox{.08ex}{\hspace{0.2em}\button{gray!30}{:}}}
	 	\button{#1}  {$\mathbf{\circlearrowleft}$}}
	 	\button{#2} {$\bumpeq$}
	 }
\newcommand{\onelabentry}[2] {\mybox{\small{\textsf{#1}}}\hspace{.25cm}\mybox{\small{\textsf{#2}}}}

\title{Open source models for the parametric study of diffraction gratings in 2D/2.5D/3D with \texttt{ONELAB}}
\author{Guillaume Demésy\thanks{Contact: \stt{guillaume.demesy@fresnel.fr}}, André Nicolet and Frédéric Zolla}

\affil[]{\small{Aix-Marseille Université, CNRS, Centrale Marseille, Institut Fresnel UMR 7249, 13013 Marseille, France}}
\date{\today}

\begin{document}
\maketitle
\chapter*{Abstract} 
This technical note aims at presenting both theoretical and practical aspects of the 
diffraction grating \stt{ONELAB} models\footnote{\url{https://gitlab.onelab.info/doc/models/wikis/Diffraction-gratings}}.

\begin{figure}[H]
  \includegraphics[draft=\flagdraft,width=\textwidth]{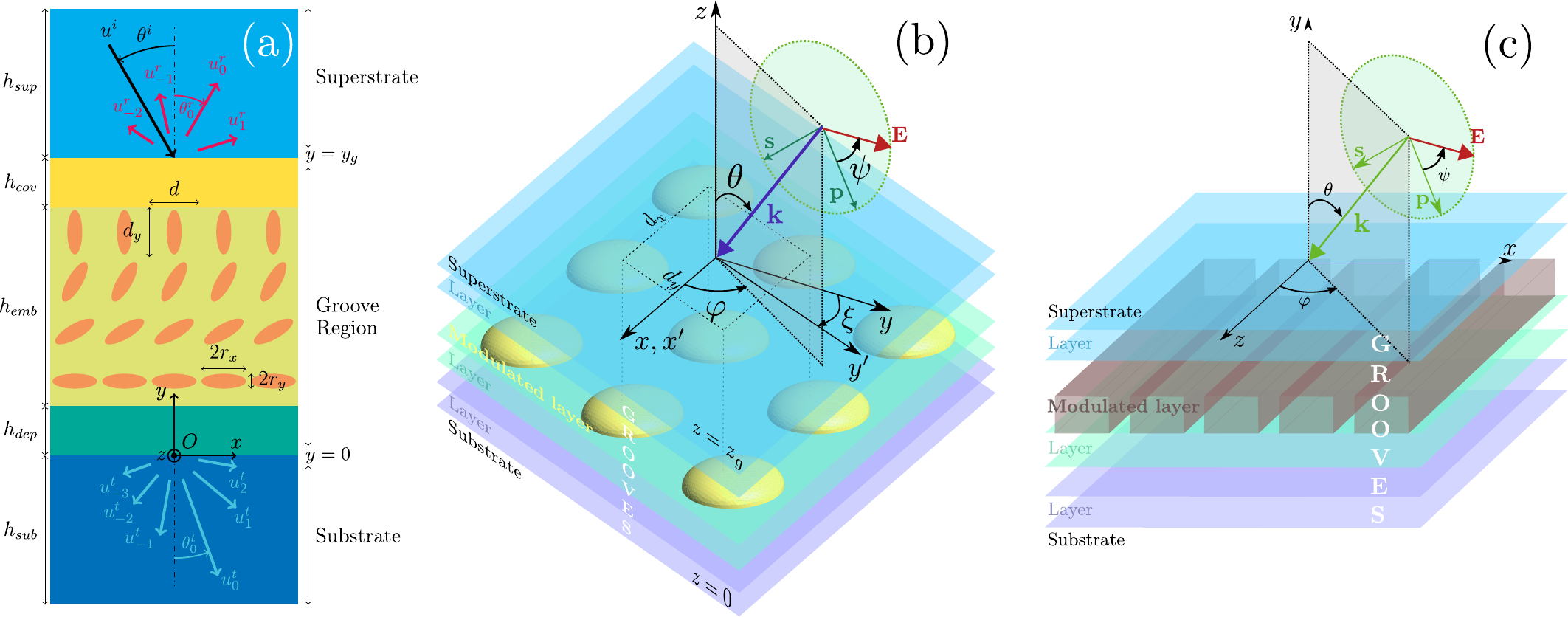}
  \caption{Scalar diffraction by a mono-dimensional grating (2D). (b) Vector diffraction by a crossed grating (3D). (c) Vector diffraction by a mono-dimensional grating (2.5D).}
  \label{fig:allgratings}
\end{figure}

The model \stt{grating2D.pro} applies to so-called mono-dimensional grating, \emph{i.e.} structures having one direction of invariance as shown in Fig.~\ref{fig:allgratings}(a). Various geometries and materials can be handled or easily added. The two classical polarization cases, denoted here $E^\parallel$ (also denoted TE in the literature)  and $H^\parallel$ (or TM), are  addressed. These are scalar problems where a scalar Helmholtz equation is solved.

The model \stt{grating3D.pro} applies to possibly skewed crossed gratings, which are 3D structures with two directions of periodicity as shown in Fig.~\ref{fig:allgratings}(b). The output of both models consist in a full energy balance of the problem computed from the field maps. This is a vector problem where a vector Helmholtz equation is solved. 

Finally, the conical incidence (2D geometry, 3D incidence, see Fig.~\ref{fig:allgratings}(c)) is treated thanks to a mixed formulation.

These models are based on free the open source pieces of software \stt{Gmsh} \cite{gmsh}, \stt{GetDP} \cite{getdp} and their interface \stt{ONELAB}. For more technical insights and a more complete bibliography, the reader is invited to refer to~\cite{demesy2007thefinite,demesy2009versatile,demesy2010allpurpose}.

\tableofcontents
\chapter{Mono-dimensional gratings: {\tt grating2D.pro}}

\section{Intro}
This chapter aims at presenting both theoretical and practical aspects regarding the 
\stt{grating\_2D} ONELAB model, mainly for educational purposes.
This model applies to so-called mono-dimensional grating, \emph{i.e.} 
structures having one direction of invariance. Various geometries and materials can be handled or easily added.
The two classical polarization cases, denoted here $E^\parallel$ (or TE) and $H^\parallel$ (or TM), are 
addressed. The output consists in a full energy balance of the problem computed from 
the field maps. For more detailed information and associated bibliography, the curious reader 
is invited refer to \cite{demesy2007thefinite}.

\begin{figure}
	\centering
	\begin{tikzpicture}
		\def\nm{0.005}
		\def\N{5}
		\def\Ny{4}
		\def\d{200*\nm}
		\def\lam{200*\nm}
		\def\dy{200*\nm}
		\def\hembed{\dy*\Ny}
		\def\hsup{600*\nm}
		\def\hsub{600*\nm}
		\def\hlayerdep{200*\nm}
		\def\hlayercov{200*\nm}
		\def\hpmltop{200*\nm}
		\def\hpmlbot{200*\nm}
		\def\wrodbot{200*\nm}
		\def\wrodtop{200*\nm}
		\def\hrod{200*\nm}
		\def\hpmltop{200*\nm}
		\def\hpmlbot{200*\nm}
    \def\xlcell{\N*\d/2+6*\d}
		\def\rx{0.9*\d/2}
		\def\ry{0.3*\dy/2}
		\def\anglerot{30}

		\def\dan{20*\nm}

		\def\Notmin{-3}
		\def\Notmax{2}
		\def\Normin{-2}
		\def\Normax{1}
		\def\thetai{30}
		\def\thetat{20}
		\def\thetardif{43}
		\def\thetatdif{30}

		\def\laxes{250*\nm}
		\def\lkin{250*\nm}
		\def\lktout{500*\nm}
		\def\lkrout{400*\nm}

    \coordinate (O) at (0,0) ;
    \coordinate (TC) at (-\hsup/2,\hlayerdep+\hembed+\hlayercov+\hsup) ;
    \coordinate (TOPCOV) at    (0,\hlayerdep+\hembed+\hlayercov) ;
    \coordinate (BC) at (\hsup/3,-\hsub) ;

    \fill[color_sub,opacity=1]  (-\N*\d/2,-\hsub            ) rectangle (\N*\d/2,0);
    \fill[color_cov,opacity=1]  (-\N*\d/2, 0                ) rectangle (\N*\d/2,\hlayerdep);
    \fill[color_embed,opacity=1](-\N*\d/2,\hlayerdep        ) rectangle (\N*\d/2,\hlayerdep+\hembed);
    \fill[color_dep,opacity=1]  (-\N*\d/2,\hlayerdep+\hembed) rectangle
																							(\N*\d/2,\hlayerdep+\hembed+\hlayercov);
    \fill[color_sup,opacity=1] (-\N*\d/2,\hlayerdep+\hembed+\hlayercov) rectangle
																							(\N*\d/2,\hlayerdep+\hembed+\hlayercov+\hsup);
		\foreach \j in {1,2,...,\Ny}{
			\foreach \i in {1,2,...,\N}{
				\fill[color_rod,
							rotate around={\anglerot*(\j-1):(-\N*\d/2+\i*\d-\d/2,\hlayerdep-\dy/2+\j*\dy)}]
								 								(-\N*\d/2+\i*\d-\d/2,\hlayerdep-\dy/2+\j*\dy)
																ellipse [x radius=\rx, y radius=\ry];};
		};
    \draw[<->]  (-\N*\d/2-\dan,-\hsub) -- (-\N*\d/2-\dan,0) node[midway,left] {$h_{sub}$};
    \draw[<->]  (-\N*\d/2-\dan,     0) -- (-\N*\d/2-\dan,\hlayerdep) node[midway,left] {$h_{dep}$};
    \draw[<->]  (-\N*\d/2-\dan,\hlayerdep) -- (-\N*\d/2-\dan,\hlayerdep+\hembed) node[midway,left] {$h_{emb}$};
    \draw[<->]  (-\N*\d/2-\dan,\hlayerdep+\hembed) -- (-\N*\d/2-\dan,\hlayerdep+\hembed+\hlayercov) node[midway,left] {$h_{cov}$};
    \draw[<->]  (-\N*\d/2-\dan,\hlayerdep+\hembed+\hlayercov) -- (-\N*\d/2-\dan,\hlayerdep+\hembed+\hlayercov+\hsup) node[midway,left] {$h_{sup}$};

    \draw[<->]  (-\d/2,\hlayerdep+\hembed+\dan) -- (\d/2,\hlayerdep+\hembed+\dan) node[midway,above] {$d$};
    \draw[<->]  (-\d/2,\hlayerdep+\hembed-\dy) -- (-\d/2,\hlayerdep+\hembed) node[midway,left] {$d_y$};
    \draw[<->]  (\d-\rx,\hlayerdep+\d/2+\ry+\dan) -- (\d+\rx,\hlayerdep+\d/2+\ry+\dan) node[midway,above] {$2r_x$};
    \draw[<->]  (\d+\rx+\dan,\hlayerdep+\d/2-\ry) -- (\d+\rx+\dan,\hlayerdep+\d/2+\ry) node[midway,right] {$2r_y$};

    \draw[<->]  (\N*\d/2+2*\dan,2*\dan) -- (\N*\d/2+2*\dan,\hlayerdep+\hembed+\hlayercov-2*\dan) node[midway,right] {Groove Region};
    \draw[<->]  (\N*\d/2+2*\dan,\hlayerdep+\hembed+\hlayercov+2*\dan) -- (\N*\d/2+2*\dan,\hlayerdep+\hembed+\hlayercov+\hsup)  node[midway,right] {Superstrate};
    \draw[<->]  (\N*\d/2+2*\dan,-\hsub) -- (\N*\d/2+2*\dan,-2*\dan)  node[midway,right] {Substrate};

		\node[right] at (\xlcell+\d,\hlayerdep+\hembed+\hlayercov+\hsup/2) {};
		\node[right] at (\xlcell+\d,-\hsub/2)         {Substrate};
		\node[right] at (\N*\d/2,0) {\small $y=0$};
		\node[right] at (\N*\d/2,\hlayerdep+\hembed+\hlayercov) {\small $y=y_g$};

    \filldraw[fill=color_sub,draw=black,opacity=.2]  (\xlcell,-\hsub-\hpmlbot   ) rectangle (\xlcell+\d,-\hsub);
    \filldraw[fill=color_sub,draw=black,opacity=1]   (\xlcell,-\hsub            ) rectangle (\xlcell+\d,0);
    \filldraw[fill=color_cov,draw=black,opacity=1]   (\xlcell, 0                ) rectangle (\xlcell+\d,\hlayerdep);
    \filldraw[fill=color_embed,draw=black,opacity=1] (\xlcell,\hlayerdep        ) rectangle (\xlcell+\d,\hlayerdep+\hembed);
    \filldraw[fill=color_dep,draw=black,opacity=1]   (\xlcell,\hlayerdep+\hembed) rectangle
																							(\xlcell+\d,\hlayerdep+\hembed+\hlayercov);
    \filldraw[fill=color_sup,draw=black,opacity=1] (\xlcell,\hlayerdep+\hembed+\hlayercov) rectangle
																							(\xlcell+\d,\hlayerdep+\hembed+\hlayercov+\hsup);
    \filldraw[fill=color_sup,draw=black,opacity=.2] (\xlcell,\hlayerdep+\hembed+\hlayercov+\hsup) rectangle
																							(\xlcell+\d,\hlayerdep+\hembed+\hlayercov+\hsup+\hpmlbot);
    \node[right] at (\xlcell+\d,-\hsub-\hpmlbot/2) {PML substrate};
    \node[right] at (\xlcell+\d,-\hsub/2)         {Substrate};
    \node[right] at (\xlcell+\d,\hlayerdep/2)     {Deposition layer};
    \node[right] at (\xlcell+\d,\hlayerdep+\hembed/2) {Embedding layer};
    \node[right] at (\xlcell+\d,\hlayerdep+\hembed+\hlayercov/2) {Cover layer};
    \node[right] at (\xlcell+\d,\hlayerdep+\hembed+\hlayercov+\hsup/2) {Superstrate};
    \node[right] at (\xlcell+\d,\hlayerdep+\hembed+\hlayercov+\hsup+\hpmltop/2) {PML Superstrate};

    \foreach \j in {1,2,...,\Ny}{
        \filldraw[fill=color_rod,draw=black,
              rotate around={\anglerot*(\j-1):(\xlcell+\d/2,\hlayerdep-\dy/2+\j*\dy)}]
                                          (\xlcell+\d/2,\hlayerdep-\dy/2+\j*\dy)
                                ellipse [x radius=\rx, y radius=\ry];
                  \node[left] at (\xlcell,\hlayerdep-\dy/2+\j*\dy) {Rod \j};
    };

    \draw [dashdotted] (TOPCOV) -- (0,\hlayerdep+\hembed+\hlayercov+2.1*\laxes) ;
    \draw [<->,thick] (0,\laxes) node[left]{$y$}
									 -- (O) node[above right]{$O$}
									 -- (\laxes,0) node[above]{$x$};
    \draw [<->,thick] (0,\laxes) node[left]{$y$}
									 -- (O) node[above right]{$O$}
									 -- (\laxes,0) node[above]{$x$};
		\draw [thick] (O) ellipse[x radius=0.1*\laxes, y radius=0.1*\laxes] node[left]{$z$};
		\fill []      (O) ellipse[x radius=0.05*\laxes, y radius=0.05*\laxes];

    \draw [->,color_inc,thick](0,\hlayerdep+\hembed+\hlayercov+1.8*\laxes) arc (90:90+\thetai:1.8*\laxes);
    \node[color_inc] at (-\laxes/2.1,\hlayerdep+\hembed+\hlayercov+2.*\laxes)  {$\theta^{i}$};
    \draw[->,color_inc,ultra thick]
							({-2.1*\lkin*sin(\thetai)},
							 { 2.1*\lkin*cos(\thetai)+\hlayerdep+\hembed+\hlayercov}) node[above]{$u^i$}
							 -- (TOPCOV);
    \draw [->,color_refl,thick](0,\hlayerdep+\hembed+\hlayercov+\laxes) arc (90:90-\thetai:\laxes);
    \node[color_refl] at (\laxes/4.8,\hlayerdep+\hembed+\hlayercov+\laxes/1.3)  {\footnotesize $\theta^{r}_0$};
    \foreach \k in {\Normin,...,\Normax}{
    	\draw[->,color_refl,ultra thick]
						({ \lktout/5*sin(\thetai+\k*\thetardif)},
						 { \lktout/5*cos(\thetai+\k*\thetardif)+\hlayerdep+\hembed+\hlayercov})
				 -- ({(\lktout/5+\lktout/(abs(\k)+2))*sin(\thetai+\k*\thetardif)},
				     {(\lktout/5+\lktout/(abs(\k)+2))  *cos(\thetai+\k*\thetardif)+\hlayerdep+\hembed+\hlayercov})
				 node[at end,above]{\small$u^r_{\k}$};

		};

    \foreach \k in {\Notmin,...,\Notmax}{
    	\draw[->,color_tran,ultra thick]
						({  \lktout/5*sin(\thetat+\k*\thetatdif)},
						 { -\lktout/5*cos(\thetat+\k*\thetatdif)})
				 -- ({( \lktout/5+\lktout/(abs(\k)+1.2))*sin(\thetat+\k*\thetatdif)},
				 		 {(-\lktout/5-\lktout/(abs(\k)+1.2))*cos(\thetat+\k*\thetatdif)})
				 node[label={[label distance=-.5cm]\thetat+\k*\thetatdif-90:\small$u^t_{\k}$ }]{};

		};
    \draw [dashdotted] (0,0) -- (0,-2.1*\laxes) ;
    \draw [->,color_tran,thick](0,-1.65*\laxes) arc (90:90+\thetat:-1.65*\laxes);
    \node[color_tran] at (280:1.8)  {\footnotesize $\theta_{0}^t$};
    \node[] at (0,-\hsub-\hpmlbot/2)  {\Large (\emph{a})};
    \node[left] at (\xlcell,-\hsub-\hpmlbot/2) {\Large (\emph{b})};

	\end{tikzpicture}
	\caption{Example of grating structure covered by the present ONELAB model.}\label{fig:grating1}
\end{figure}

\section{Theoretical model}
\subsection{Set up of the problem and notations}
We denote by $\x$, $\y$ and $\z$, the unit vectors of the axes of an
orthogonal co-ordinate system $Oxyz$. Time-harmonic regime is assumed; 
consequently, the electric and magnetic fields
are represented by the complex vector fields $\bE$ and $\bH$ with a
time dependence chosen in $\exp(-i\omega t)$. We are now considering 
2D structures $0z$ is the axis of invariance.

Besides, in this model, we assume that the tensor fields of relative
permittivity $\tensepsr$ and relative permeability $\tensmur$ can be
written as follows:
\begin{equation}\label{eq:tenseps}
\tensepsr=%
\left( %
\begin{array}{ccc}
  \varepsilon_{xx} & \bar{\varepsilon}_{a} & 0 \\
  \varepsilon_{a} & \varepsilon_{yy} & 0 \\
  0 & 0 & \varepsilon_{zz}
\end{array}
\right) %
\quad \hbox{and} \quad
\tensmur=%
\left( %
\begin{array}{ccc}
  \mu_{xx} & \bar{\mu}_{a} & 0 \\
  \mu_{a} & \mu_{yy} & 0 \\
  0 & 0 & \mu_{zz}
\end{array}
\right) \; , %
\end{equation}
where $\varepsilon_{xx},\varepsilon_{a},\dots \mu_{zz}$ are possibly
complex valued functions of the two variables $x$ and $y$ and where
$\bar{\varepsilon}_{a}$ (resp. $\bar{\mu}_{a}$) represents the
conjugate complex of
$\varepsilon_{a}$ (resp. $\mu_{a}$). \textit{%
These kinds of materials are said to be $z$--anisotropic}. It is of
importance to note that with such tensor fields, lossy materials can
be studied (the lossless materials correspond to tensors with real
diagonal terms represented by Hermitian matrices) and that the
problem is invariant along the $z$--axis but the tensor fields can
vary continuously (gradient index gratings) or discontinuously (step
index gratings). We define the wavenumber $k_0 := \omega/c$.

The gratings that we are dealing with are made of three regions (See
Fig.~\ref{fig:grating1}a).
\begin{itemize}
  \item
  \textit{The superstrate} ($y>y_g$) which is supposed to be
  homogeneous, isotropic and lossless and characterized solely by its
  real valued relative permittivity $\varepsilon_r^+$ 
  and its relative permeability
  $\mu_r^+$. We denote $k^+:=k_0\, \sqrt{\varepsilon_r^+ \mu_r^+}$.
  \item
  \textit{The substrate} ($y<0$) is supposed to be homogeneous
  and isotropic and therefore characterized by its relative
  permittivity $\varepsilon_r^-$ and its relative permeability
  $\mu_r^-$. We denote $k^-:=k_0\, \sqrt{\varepsilon_r^- \mu_r^-}$.
  \item
  \textit{The groove region} ($0<y<y_g$) is heterogeneous and
  $z$--anisotropic. It is characterized by the two tensor fields
  $\tensepsr^g(x,y)$ and $\tensmur^g(x,y)$. It is worth noting that the
  method presented in this paper does work irrespective of whether the
  tensor fields are piecewise constant. The grating periodicity along 
  $x$--axis will be denoted $d$.
\end{itemize}

This grating is illuminated by an incident plane wave of wave vector
$$\bk_\shortdownarrow^+=\alpha\, \x \textcolor{black}{-} \beta^+ \, \y=k^+ \left(\sin \theta_0 \x -
 \cos \theta_0 \y \right),$$ whose electric field ($E^\parallel$ polarization case ) (
 resp. magnetic field ($H^\parallel$ )) is linearly polarized along the $z$--axis:
\begin{equation}\label{eq:incPW}
\bE_e^0=\mathbf{A}_e^0 \, \exp(i \bk_\shortdownarrow^+ \cdot \br)\, \z \quad
(\hbox{resp. $\bH_m^0=\mathbf{A}_m^0 \, \exp(i \bk_\shortdownarrow^+ \cdot \br)\,
\z$}) \; ,
\end{equation}
where $\mathbf{A}_e^0$ (resp. $\mathbf{A}_m^0$) is an arbitrary
complex number. The magnetic (resp. electric) field derived from
$\bE_e^0$ (resp. $\bH_m^0$) is denoted $\bH_e^0$ (resp. $\bE_m^0$)
and the electromagnetic field associated with the incident field is
therefore denoted ($\bE^0,\bH^0$) which is equal to
($\bE_e^0,\bH_e^0$) (resp. ($\bE_m^0,\bH_m^0$)).

The problem of diffraction that we address in this paper is
therefore to find Maxwell's equation solutions in harmonic regime
\textit{i.e.} the unique solution ($\bE,\bH$) of:
\begin{subequations}\label{eq:Maxwell}
\begin{numcases}{}
\curl\, \bE=\textcolor{black}{+}i \omega \mu_0 \, \tensmur \bH \label{eq:MaxwellrotE}\\
\curl\, \bH=\textcolor{black}{-}i \omega \varepsilon_0 \, \tensepsr \bE
\label{eq:MaxwellrotH}
\end{numcases}
\end{subequations}
such that the diffracted field satisfies an \textit{Outgoing Waves
Condition} (O.W.C. \cite{petit_ondes_1992,petit1980electromagnetic}) and where $\bE$ and $\bH$
are quasi-periodic functions with respect to the $x$ co-ordinate.

\subsection{Appropriate diffracted field formulation}
\subsubsection{Decoupling of fields and $z$--anisotropy}%
We assume that $\tensdelta(x,y)$ is a $z$--anisotropic tensor field
($\delta_{xz}=\delta_{yz}=\delta_{zx}=\delta_{zy}=0$). Moreover, the
left upper matrix extracted from $\tensdelta$  is denoted
$\tensdeltat$, namely:
\begin{equation}
\tensdeltat=%
\left( %
\begin{array}{cc}
  \delta_{xx} & \bar{\delta}_{a}  \\
  \delta_{a} & \delta_{yy}
\end{array}
\right) %
 \; . %
\end{equation}
For $z$--anisotropic materials, with non-conical incidence, the problem
of diffraction can be split into two fundamental cases ($H^\parallel$ case and
$E^\parallel$ case). This property results from the following equality which
can be easily derived:%
\begin{equation}
- \curl \left ( {\tensdelta}^{-1} \curl \left( u \z \right)\right)=
\div \left( \frac{\tensdeltat^T}{\det(\tensdeltat)} \grad u \right) \z
\; ,
\end{equation}
where $u$ is a function which does not depend on the $z$ variable.
From the previous equality, it appears that the non-conical problem of
diffraction amounts to looking for an
electric (resp. magnetic) field which is polarized along the
$z$--axis ; $\bE=e(x,y)\, \z$ (resp. $\bH=h(x,y)\, \z$). The
functions $e$ and $h$ are therefore solutions of similar
differential equations:
\begin{equation}\label{eq:defL}
\mathscr{L}_{\tensxi, \chi}(u):=\div\left( \tensxi \, \grad u
\right) + k_0^2 \chi\, u =0
\end{equation}
with
\begin{equation}\label{eq:defxiTE}
u=e, \quad \tensxi= \tensmut^T/\det(\tensmut), \quad
\chi=\varepsilon_{zz} \; ,
\end{equation}
in the $E^\parallel$ case and
\begin{equation}\label{eq:defxiTM}
u=h, \quad \tensxi= \tensepst^T/\det(\tensepst), \quad \chi=\mu_{zz}
\; ,
\end{equation}
in the $H^\parallel$ case.

\subsubsection{Reducing the diffraction problem to a radiation problem with localized sources}
In its initial form, the problem of diffraction summed up by Eq.~(\ref{eq:defL})
is not well suited to the Finite Element Method. We propose to split the unknown
function $u$ into a sum of two functions $u_1$ and $u_2^d$, the
first term being known as a closed form and the latter being a
solution of a radiation problem whose sources are localized
within the obstacles. This is, in essence, a diffracted field formulation
extended to the case where the substrate and superstrate are made 
of different materials.

We have assumed that outside the groove
region (cf. Fig. \ref{fig:grating1}), the tensor field
$\tensxi$ and the function $\chi$ are constant and equal
respectively to $\tensxi^-$ and $\chi^-$ in the substrate ($y<0$)
and equal respectively to $\tensxi^+$ and $\chi^+$ in the
superstrate ($y>y_g$). Besides, for the sake of clarity, the
superstrate is supposed to be made of an isotropic and lossless
material and is therefore solely defined by its relative
permittivity $\varepsilon_r^+$ and its relative permeability $\mu_r^+$,
which leads to:
\begin{equation}\label{eq:defplusTE}
\tensxi^+= \frac{1}{\mu_r^+} \, \mathrm{Id}_2 \quad \hbox{and} \quad
\chi^+=\varepsilon_r^+ \quad \hbox{in $H^\parallel$ case}
\end{equation}
or
\begin{equation}\label{eq:defplusTM}
\tensxi^+= \frac{1}{\varepsilon_r^+} \, \mathrm{Id}_2 \quad \hbox{and}
\quad \chi^+=\mu_r^+ \quad \hbox{in $E^\parallel$ case,}
\end{equation}
where $\mathrm{Id}_2$ is the $2\times 2$ identity matrix. With such
notations, $\tensxi$ and $\chi$ are therefore defined as follows:
\begin{equation}\label{eq:def_xi_chi_global}
\tensxi(x,y):= \left \{
\begin{array}{lcc}
  \tensxi^+ & \hbox{for} & y>y_g \\
 \tensxi^g(x,y) & \hbox{for} & y_g>y>0\\
  \tensxi^- & \hbox{for} & y<0
\end{array}
\right .%
\; , \; \chi(x,y):= \left \{
\begin{array}{lcc}
  \chi^+ & \hbox{for} & y>y_g \\
 \chi^g(x,y) & \hbox{for} & y_g>y>0\\
  \chi^- & \hbox{for} & y<0 \; .
\end{array}
\right .
\end{equation}
It is now apropos to introduce an auxiliary tensor field $\tensxi_1$
and an auxiliary function $\chi_1$:
\begin{equation}
\tensxi_1(x,y):= \left \{
\begin{array}{ccc}
  \tensxi^+ & \hbox{for} & y>0 \\
  \tensxi^- & \hbox{for} & y<0
\end{array}
\right .%
\; , \; \chi_1(x,y):= \left \{
\begin{array}{ccc}
  \chi^+ & \hbox{for} & y>0 \\
  \chi^- & \hbox{for} & y<0 \; ,
\end{array}
\right .
\end{equation}
these quantities corresponding, of course, to a simple plane
interface. Besides, we introduce the constant tensor field
$\tensxi_0$ which is equal to $\tensxi^+$ everywhere and a constant
scalar field $\chi_0$ which is equal to $\chi^+$ everywhere.
Finally, we denote $u_0$ the function which equals the incident
field $u^{\mathrm{inc}}$ in the superstrate and vanishes elsewhere:
\begin{equation}
u_0(x,y):= \left \{
\begin{array}{ccc}
  u^{\mathrm{inc}} & \hbox{for} & y>y_g \\
  0 & \hbox{for} & y<y_g
\end{array}
\right .%
\end{equation}

We are now in a position to reformulate the diffraction problem 
of interest. The function $u$ is the unique solution of
\begin{equation}\label{eq:defLOWC}
\mathscr{L}_{\tensxi, \chi}(u)=0 \quad \hbox{such that $u^d:=u-u_0$
satisfies an O.W.C.}
\end{equation}
In order to reduce this problem of diffraction to a radiation
problem, an intermediate function is necessary. This function,
called $u_1$, is defined as the unique solution of the equation:
\begin{equation}\label{eq:defL1}
\mathscr{L}_{\tensxi_1, \chi_1}(u_1)=0 \quad \hbox{such that
$u_1^d:=u_1-u_0$ satisfies an O.W.C.}
\end{equation}

The function $u_1$ corresponds thus to \textit{an annex problem}
associated to a simple interface and can be solved in closed form
and \textit{from now on is considered as a known function}. As
written above, we need the function $u_2^d$ which is simply defined
as the difference between $u$ and $u_1$:
\begin{equation}\label{eq:u2d}
u_2^d := u - u_1 =u^d-u_1^d\; .
\end{equation}
The presence of the superscript $d$ is, of course, not irrelevant :
As the difference of two diffracted fields, the O.W.C. of $u_2^d$ is
guaranteed (which is of prime importance when dealing with PML cf.
\ref{subsec:PML}). As a result, the Eq.~(\ref{eq:defLOWC}) becomes:
\begin{equation}\label{eq:radiating}
\mathscr{L}_{\tensxi, \chi}(u_2^d)=-\mathscr{L}_{\tensxi, \chi}(u_1)
\; ,
\end{equation}
where the right hand member is a scalar function which may be
interpreted as a \textit{known source term} $-\mathscr{S}_1(x,y)$
\textit{and the support of this source is localized only within the
groove region}. To prove it, all we have to do is to use Eq.
(\ref{eq:defL1}):
\begin{equation}
\mathscr{S}_1:=\mathscr{L}_{\tensxi,
\chi}(u_1)=\mathscr{L}_{\tensxi,
\chi}(u_1)-\underbrace{\mathscr{L}_{\tensxi_1,
\chi_1}(u_1)}_{=0}=\mathscr{L}_{\tensxi-\tensxi_1, \chi -
\chi_1}(u_1) \; .
\end{equation}
Now, let us point out that the tensor fields $\tensxi$ and
$\tensxi_1$ are identical outside the groove region and the same
holds for $\chi$ and $\chi_1$. The support of $\mathscr{S}_1$ is
thus localized within the groove region as expected. It remains to
compute more explicitly the source term $\mathscr{S}_1$. Making use
of the linearity of the operator $\mathscr{L}$ and the equality
$u_1=u_1^d+u_0$, the source term can be split into two terms:
\begin{equation}\label{eq:defsource}
\mathscr{S}_1=\mathscr{S}_1^0+\mathscr{S}_1^d \; ,
\end{equation}
where
\begin{equation}\label{eq:defsource0}
\mathscr{S}_1^0=\mathscr{L}_{\tensxi-\tensxi_1, \chi - \chi_1}(u_0)
\end{equation}
and
\begin{equation}\label{eq:defsourced}
\mathscr{S}_1^d=\mathscr{L}_{\tensxi-\tensxi_1, \chi -
\chi_1}(u_1^d) \; .
\end{equation}
Now, since $u_0$ is nothing but a plane wave $u_0= \exp(i \bk_\shortdownarrow^+ \cdot \rr)$  (with $\bk_\shortdownarrow^+=\alpha\,\x \textcolor{black}{-} \beta^+\,\y$), it is sufficient to give $\grad u_0=i \bk_\shortdownarrow^+ \, u_0$  for the weak formulation associated with Eq.~(\ref{eq:radiating}):
\begin{equation}\label{eq:dev_source0}
\mathscr{S}_1^0= \left\{i \div \left[ \left( \tensxi^+ -
\tensxi\right) \bk_\shortdownarrow^+ \, \exp(\bk_\shortdownarrow^+ \cdot \rr) \right] + k_0^2\left(
\chi^+- \chi\right)\exp(i \bk_\shortdownarrow^+ \cdot \rr) \right\}\; .
\end{equation}
The same holds for the term associated with the diffracted field
($u_1^d=\rho\,\exp(i \bk_\shortuparrow^+ \cdot \rr)$, with
($\bk_\shortuparrow^+=\alpha\,\x \textcolor{black}{+} \beta^{\textcolor{black}{+}}\,\y$)):
\begin{equation}\label{eq:dev_source1}
\mathscr{S}_1^d= \rho \left\{i \div \left[  \left( \tensxi^+ -
\tensxi\right) \bk_\shortuparrow^+ \, \exp(i \bk_\shortuparrow^+ \cdot \rr) \right] + k_0^2\left(
\chi^+- \chi\right)\exp(i \bk_\shortuparrow^+ \cdot \rr) \right\}\; ,
\end{equation}
where $\rho$ is nothing but the complex reflection coefficient associated with the simple interface :
\begin{equation}\label{eq:coeff_reflexion}
\rho = \frac{p^+-p^-}{p^++p^-} \; \hbox{with}\;p^\pm\,=\, \left \{
\begin{array}{ccc}
    \;\beta^\pm\;\hbox{in the $E^\parallel$ case}\\
    \\
    \;\frac{\beta^\pm}{\varepsilon_r^\pm}\;\hbox{in the $H^\parallel$ case}\\
\end{array}
\right .
\end{equation}

\subsubsection{An important remark about the choice of the unknown diffracted field}
It is important to understand that we have several choices for the unknown field. Our goal is to formulate an equivalent problem for which the support of the sources is bounded and inside the computational domain.
\begin{figure}[H]
	\centering
	\includegraphics[width=.99\textwidth]{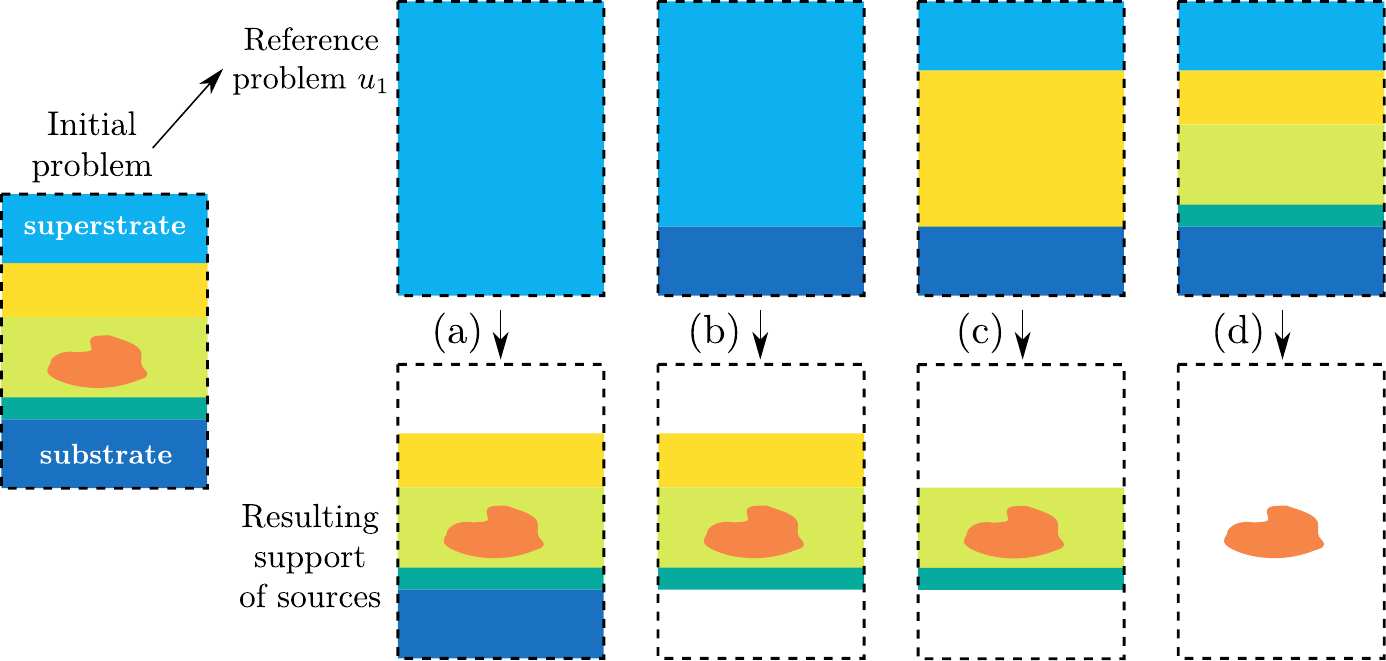}
	\caption{Some possible choices for the annex problem.}
	\label{fig:locsources}
\end{figure}
Figure~\ref{fig:locsources} illustrates the impact of the choice of the annex problem upon the support of the sources. With choice (a), the annex problem is nothing but the response of freespace to the desired incident field, which is trivial to compute indeed. However, the corresponding scattering problem (see bottom inset) has now sources in the substrate which is unbounded. So we didn't gain much here: Instead of solving for the total field with sources inside the unbounded superstrate, we now have to solve a scattered field with sources in the unbounded substrate. We understand here the importance of considering a total field and an annex field satisfying the same radiation condition both inside the substrate and the superstrate. Choices (b,c,d) in Figure~\ref{fig:locsources} are more suitable candidates since they take into account the impedance mismatch between the superstrate and the substrate. Choice~(b) amounts to nothing but compute the Fresnel coefficients of the planar interface. For choices~(c,d), one has to compute first the  response of a multilayer.

When considering periodic structures, the simplest choice is choice (b) and this is the choice made throughout this document. However, note that when considering isolated scatterers embedded in a multilayer, we need to extend the discussion to consider the \enquote{lateral} radiation condition in order to avoid incoming sources from an infinite distance along $x$ in the scattering problem. The only choice is then choice (d), where one has to pre-compute the field scattered by the 1D invariant multilayered stack.

\subsubsection{Quasi-periodicity and weak formulation}
The weak formulation follows the classical lines and is based on the construction of a weighted residual of Eq.~(\ref{eq:defL}), which is multiplied by the complex conjugate of a weight function $u'$ and integrated by part to obtain :

\begin{eqnarray}
\mathscr{R}_{\tensxi, \chi}(u,u') = \int_\Omega -\left(\tensxi \,
\grad u \right) \cdot \overline{\grad u'} + k_0^2\,\chi\, u  \;
\overline{u'} \d\Omega + \int_{\partial \Omega} \overline{u'}
\left( \tensxi \, \grad u \right) \cdot \mathbf{n}\d\ell
\end{eqnarray}
The solution $u$ of the weak formulation can therefore be defined as
the element of the space $H^1(\grad,d,\alpha)$ of quasi-periodic functions
(i.e. such that $u(x,y)=u_\#(x,y)e^{ikx}$ with
$u_\#(x,y)=u_\#(x+d,y)$, a $d$-periodic function and where both $u$ and $\grad\,u$ are square integrable) such that: 
\begin{equation} \mathscr{R}_{\tensxi,
\chi}(u,u')=0 \; \; \forall u' \in H^1(\grad,d,\alpha). 
\end{equation} 
As for the boundary term introduced by the integration by part,  it can be classically set to zero at the PML endings by imposing Dirichlet conditions on a part of the boundary (the value of $u$ is imposed and the weight function $u'$ can be chosen equal to zero on this part of the boundary) or by imposing homogeneous Neumann conditions $(\tensxi \grad u) \cdot \mathbf{n}=0$ on another part of the boundary (and $u$ is therefore an unknown to be determined on the boundary). A third possibility (applied here to lateral boundaries) are the so-called quasi-periodicity conditions of particular importance in the modeling of gratings. Denote by $\Gamma_l$ and $\Gamma_r$ the lines parallel to the $y$--axis delimiting a cell of the grating respectively from its left and right neighbor cell. Considering that both  $u$ and $u'$ are in $L^2(\curl,d,k)$, the boundary term for $\Gamma_l \cup  \Gamma_r$ is

 $$\int_{\Gamma_l \cup \Gamma_r} \overline{u'} \left( \tensxi \, \grad
u \right) \cdot \mathbf{n} dS= \int_{\Gamma_l \cup \Gamma_r}
\overline{u'_\#}e^{-ikx} \left( \tensxi \, \grad (u_\#e^{+ikx})
\right) \cdot \mathbf{n} dS =$$
$$ \int_{\Gamma_l \cup \Gamma_r}
\overline{u'_\#} \left( \tensxi \,( \grad u_\# +i k u_\#
\mathbf{x})\right) \cdot \mathbf{n} dS = 0$$ because the integrand
$\overline{u'_\#} \left( \tensxi \,( \grad u_\# +i k u_\#
\mathbf{x})\right) \cdot \mathbf{n}$ is periodic along $x$ and the
normal $\mathbf{n}$ has opposite directions on $\Gamma_l$ and
$\Gamma_r$ so that the contributions of these two boundaries have
the same absolute value with opposite signs. The contribution of the
boundary terms vanishes therefore naturally in the case of
quasi-periodicity.

The finite element method is based on this weak formulation and both
the solution and the weight functions are classically chosen in a
discrete space made of linear or quadratic Lagrange elements, i.e.
piecewise first or second order two variable polynomial
interpolation built on a triangular mesh of the domain $\Omega$ (cf.
Fig.\ref{fig:grating1}b). Dirichlet and Neumann conditions
may be used to truncate the PML domain in a region where the field
(transformed by the PML) is negligible. The quasi-periodic boundary
conditions are imposed by considering the $u$ as unknown on
$\Gamma_l$ (in a way similar to the homogeneous Neumann condition
case) while, on $\Gamma_r$, $u$ is forced equal to the value of the
corresponding point on $\Gamma_l$ (i.e. shifted by a quantity $-d$
along $x$) up to the factor $e^{i\,\alpha\,d}$. 
The practical implementation in the finite element method is described in details in
Ref.~\cite{renversez2012foundations}.

\subsubsection{Perfectly Matched Layers}\label{subsec:PML}
The main drawback encountered in electromagnetism when tackling theory of gratings through the finite element method is the non-decreasing behaviour of the propagating modes in superstrate and substrate (if those are made of lossless materials): The PML has been introduced by 
{berenger94perfec-match-layer} in order to get round this obstacle. Standard PMLs constant profile are implemented in the present model.

\subsubsection{Post-processing: Diffraction efficiencies calculation}\label{subsection:diff_eff}
The rough result of the FEM calculation is the total complex field solution of Eq.~(\ref{eq:defL}) at each point of the bounded domain. We deduce from $u^d$ (cf Eq.~(\ref{eq:defLOWC})) the diffraction efficiencies with the following method. The superscripts $ ^+$ (resp. $ ^-$) correspond to quantities defined in the superstrate (resp. substrate) as previously.

On the one hand, since $u^d$ is quasi-periodic along the $x$--axis, it can be expanded as a Rayleigh expansion (see for instance \cite{petit_ondes_1992,petit1980electromagnetic}):
\begin{equation}\label{eq:devfourier}
\hbox{for}\;y<0\;\hbox{and}\;y>y_g,\;u^{d}(x,y) = \sum_{n \in
\mathbb{Z}}\,u^d_n(y)\,e^{i\alpha_n x}\; ,
\end{equation}
where
\begin{equation}\label{eq:coefffourier1}
u^d_n(y)= \frac{1}{d}\int_{-d/2}^{d/2}u^d(x,y)e^{-i\alpha_n x}dx
 \;\; \hbox{with} \;\; \alpha_n = \alpha+\frac{2\pi}{d}n\,\,.
\end{equation}
On the other hand, introducing Eq.~(\ref{eq:devfourier}) into Eq.~(\ref{eq:defL}) leads to the Rayleigh coefficients :
\begin{equation}\label{eq:coefffourier2.a}
u^d_n(y)= \left \{
\begin{array}{ccc}
  s_n\,e^{+i\beta_n^+ y} + r_n\,e^{-i\beta_n^+ y} & \hbox{for} & y>y_g \\
  \\
  u_n\,e^{-i\beta_n^- y} + t_n\,e^{+i\beta_n^- y} & \hbox{for} & y<0 \\
\end{array}
\right . \; \hbox{with} \; \beta_{n}^{\pm^2} = k^{\pm^2} -
\alpha_n^2
\end{equation}
For a temporal dependence  in $e^{+i\omega t}$ , the O.W.C. imposes $s_n=u_n=0$. Combining Eq.~(\ref{eq:coefffourier1}) and Eq.~(\ref{eq:coefffourier2.a}) at a fixed $y_0$ altitude leads to:
\begin{equation}\label{eq:AnBn}
\left \{
\begin{array}{llll}
  r_n &= \displaystyle\frac{1}{d}\displaystyle\int_{-d/2}^{d/2}u^d(x,y_0)\,e^{-i(\alpha_n x-\beta_n^+ y_0)}\,\mathrm{d}x & \hbox{for} & y_0>y_g\\
    \\
  t_n &= \displaystyle\frac{1}{d}\displaystyle\int_{-d/2}^{d/2}u^d(x,y_0)\,e^{-i(\alpha_n x+\beta_n^- y_0)}\,\mathrm{d}x & \hbox{for} & y_0<0\\
\end{array}
\right .\,.
\end{equation}
We extract these two coefficients by numerical integration along $x$ from a cut of the previously calculated field map at altitudes $y_0=-h_{sub}$ in the substrate and $y_0=y_g+h_{sup}$ in the superstrate. From this we immediately deduce the reflected and transmitted diffracted efficiencies of propagative orders ($T_n$ and $R_n$) defined by :
\begin{equation}\label{eq:diffefficiency}
\left \{
\begin{array}{llll}
    R_n &:=\,r_n\,\overline{r_n}\,\displaystyle\frac{\beta_n^+}{\beta^+} & \hbox{for} & y_0>y_g\\
    \\
    T_n &:=\,t_n\,\overline{t_n}\,\displaystyle\frac{\beta_n^-}{\beta^-}\,\displaystyle\frac{\gamma^+}{\gamma^-} & \hbox{for} & y_0<0\\
\end{array}
\right . \hbox{with}\;\gamma^\pm\,=\, \left \{
\begin{array}{ccc}
    1\;\hbox{in the $E^\parallel$ case}\\
    \\
    \;\varepsilon_r^\pm\;\hbox{in the $H^\parallel$ case}\\
\end{array}
\right . \, .
\end{equation}

\section{{\ttfamily ONELAB} model description}
In this section, the parameters of the \stt{ONELAB} model are briefly commented in their order of appearance in the \stt{gmsh}'s left panel.
\subsection{{\ttfamily Geometry}}
	\subsubsection{{\ttfamily Grating period}}
		\begin{itemize}
			\item[$\bullet$] \onelabentry{value}{grating period [nm]} allows to set the period $d$  of the grating given in nanometers.
		\end{itemize}
	\subsubsection{{\ttfamily Stack thicknesses}}
		The following parameters can take any positive float value.
		\begin{itemize}
			\item[$\bullet$] \onelabentry{value}{substrate thickness [nm]}
			allows to set $h_{subs}$, given in nanometers. Quantitative results should not depend on this parameter since the substrate is by definition a half plane.
			\item[$\bullet$] \onelabentry{value}{deposit layer thickness [nm]}
			allows to set $h_{dep}$, given in nanometers.
			\item[$\bullet$] \onelabentry{value}{cover layer thickness [nm]}
			allows to set $h_{cov}$, given in nanometers.
			\item[$\bullet$] \onelabentry{value}{superstrate thickness [nm]}
			allows to set $h_{sup}$, given in nanometers. Quantitative results should not depend on this parameter since the superstrate is by definition a half plane.
		\end{itemize}
		Note that $h_{emb}$ is set by the diffractive element dimensions detailed hereafter.
	\subsubsection{{\ttfamily Diffractive element dimensions}}
\begin{figure}[H]
			\centering
				\includegraphics[draft=\flagdraft,width=.9\textwidth]{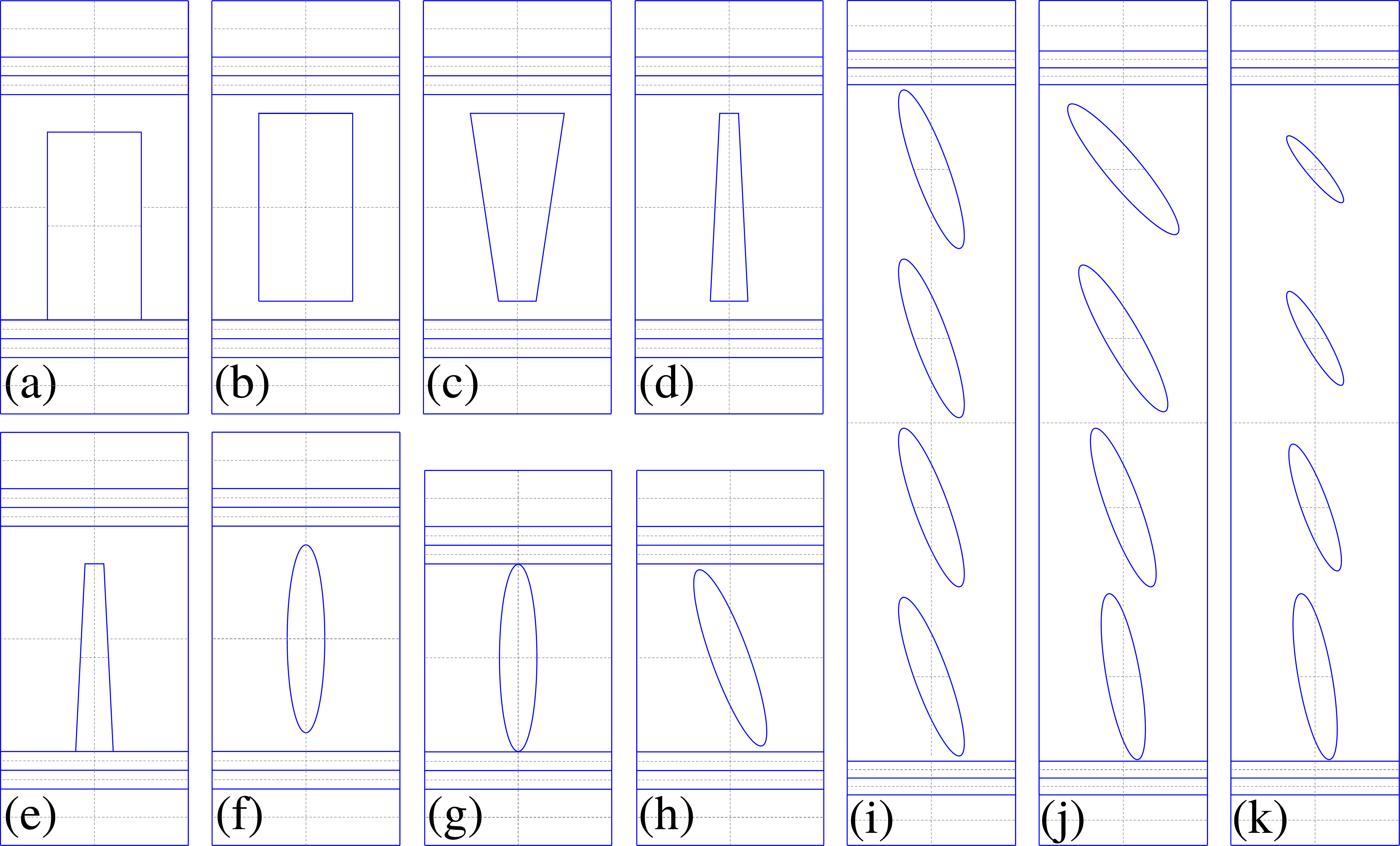}
					\caption{
						(a) \footnotesize{Initial configuration described in Fig.~\ref{fig:lamellar}.}
						(b) \footnotesize{Unchecking \onelabentry{\checkboxC}{glue rod to substrate?}.}
						(c) \footnotesize{Decreasing value of \onelabentry{value}{bottom rod width [nm]}.}
						(d) \footnotesize{Decreasing value of \onelabentry{value}{top rod width [nm]}.}
						(e) \footnotesize{Checking back \onelabentry{\checkboxC}{glue rod to substrate?}.}
						(f) \footnotesize{Choosing elliptical section in \onelabentry{\DOWNarrow\,\,menu}{rod section shape}.}
						(g) \footnotesize{Decreasing value of \onelabentry{value}{embedding layer thickness or\dots}.}
						(h) \footnotesize{Increasing value of \onelabentry{value}{rotate rod [deg]}.}
						(i) \footnotesize{Increasing value of \onelabentry{value}{number of rods [-]}.}
						(j) \footnotesize{Checking \onelabentry{value}{chirp angle?}.}
						(k) \footnotesize{Checking \onelabentry{value}{chirp size ?} and decreasing \onelabentry{value}{chirp size factor [\%]}.}
					}
					\label{fig:diff_element1}
		\end{figure}

	In order to illustrate the various grating or photonic crystal slabs covered by this model, let us start from the lamellar grating situation shown in
		Fig.~\ref{fig:lamellar}:
		\begin{itemize}
			\item[$\bullet$]  \onelabentry{\checkboxC}{glue rod to substrate?}
												having the element
												relying directly on the substrate changes the topology,
												it needs to be specified. The checking/unchecking of
												this box is illustrated in Figs.~\ref{fig:diff_element1}(a-b) and (d-e).
			\item[$\bullet$] \onelabentry{\DOWNarrow\,\,menu}{rod section shape}
												Choose between elliptical or trapezoidal rod section.
												See Figs.~\ref{fig:diff_element1}(e-f).
			\item[$\bullet$] \onelabentry{value}{number of rods [-]}
												Integer value setting the number of rods to consider along $y$ axis 
												spaced by $d_y$ (see below).
												See Figs.~\ref{fig:diff_element1}(h-i).
			\item[$\bullet$] \onelabentry{value}{bottom rod width [nm]}
												In case of a trapezoidal rod,
												this value sets the bottom width Figs.~\ref{fig:diff_element1}(c).
												In case of an elliptical rod, this value has no effect.
												See Figs.~\ref{fig:diff_element1}(b-c).
			\item[$\bullet$] \onelabentry{value}{top rod width [nm]}
												In case of a trapezoidal rod,
												this value sets the top width Figs.~\ref{fig:diff_element1}(c).
												In case of an elliptical rod, this value sets its diameter ($2r_x$).
												See Figs.~\ref{fig:diff_element1}(c-d).
			\item[$\bullet$] \onelabentry{value}{rod thickness [nm]}
												In case of a trapezoidal rod,
												this value sets the thickness (dimension along $y$).
												In case of an elliptical rod, this value sets its diameter ($2r_y$).
			\item[$\bullet$] \onelabentry{value}{embedding layer thickness or
																\enquote{period} along y if number of rods >1, [nm]}
											 If the number of rods is set to 1, this value sets $h_{emb}$.
											 In case of several rods, this value sets their periodic spacing 
											 along $y$ ($d_y$). See Figs.~\ref{fig:diff_element1}(i).
			\item[$\bullet$] \onelabentry{value}{rotate rod [deg]}
											 Rotates the rod around himself (axis formed by its barycenter, the $Oz$
											 direction). See Figs.~\ref{fig:diff_element1}(g-h).
			\item[$\bullet$] \onelabentry{value}{chirp angle?}
			 									In case of several rods, the rotation angle of the next
												rod along increasing values of $y$ is increased by the value described
												in the previous item.
												See Figs.~\ref{fig:diff_element1}(i-j).
			\item[$\bullet$] \onelabentry{value}{chirp size factor [\%]}
			\item[$\bullet$] \onelabentry{\checkboxC}{chirp size?}
			 									In case of several rods, the size of the next
												rod along increasing values of $y$ is scaled of the value given
												in the previous item.
												See Figs.~\ref{fig:diff_element1}(j-k).
		\end{itemize}
\subsection{{\ttfamily Materials}}
	\subsubsection{{\ttfamily Dispersive materials}}
		For each constitutive layer, a choice of materials is proposed.
		The file \stt{grating\_2D\_materials.pro} contains frequency
		dispersion tables for some selected materials.
		Relative permittivity values are linearly interpolated
		using these tables. The available materials are currently:
		\begin{itemize}
			\item[$\bullet$] \mybox{\textsf{Air}}          : freespace
			\item[$\bullet$] \mybox{\textsf{SiO2}}         : silicon dioxide
			\item[$\bullet$] \mybox{\textsf{Ag (palik)}}   : silver, values from Ref.~\cite{palik1998handbook}
			\item[$\bullet$] \mybox{\textsf{Al (palik)}}   : aluminium, values from Ref.~\cite{palik1998handbook}
			\item[$\bullet$] \mybox{\textsf{Au (johnson)}} : gold, values from Ref.~\cite{johnson_optical_1972}
			\item[$\bullet$] \mybox{\textsf{Nb2O5}}        : niobium pentoxide, values from Ref.~\cite{refractiveindex}
			\item[$\bullet$] \mybox{\textsf{ZnSe}}         : zinc selenide, values from Ref.~\cite{refractiveindex}
			\item[$\bullet$] \mybox{\textsf{MgF2}}         : magnesium fluoride, values from Ref.~\cite{refractiveindex}
			\item[$\bullet$] \mybox{\textsf{TiO2}}         : titanium dioxide, values from Ref.~\cite{refractiveindex}
			\item[$\bullet$] \mybox{\textsf{PMMA}}         : methyl polymethacrylate, values from Ref.~\cite{refractiveindex}
			\item[$\bullet$] \mybox{\textsf{Si}}           : silicon, values from Ref.~\cite{palik1998handbook}
			\item[$\bullet$] \mybox{\textsf{ITO}}          : indium tin oxide, values from Ref.~\cite{refractiveindex}
			\item[$\bullet$] \mybox{\textsf{Cu (palik)}}   : copper, values from Ref.~\cite{palik1998handbook}
			\item[$\bullet$] \mybox{\textsf{custom 1}}     : custom dispersion free material, see next section
			\item[$\bullet$] \mybox{\textsf{custom 2}}     : custom dispersion free material, see next section
			\item[$\bullet$] \mybox{\textsf{custom 3}}     : custom dispersion free material, see next section
		\end{itemize}
		It is easy to add another material, instructions are given in comments at the beginning of
		the file \stt{grating\_2D\_materials.pro}.
	\subsubsection{{\ttfamily Custom non-dispersive materials}}
		Another possibility is to set a material permittivity to \mybox{\textsf{custom 1}},
		\mybox{\textsf{custom 2}}, or \mybox{\textsf{custom 3}}
		in which case the real and imaginary parts of the complex relative permittivity
		will be set to the one manually specified in this section. Beware that due to the $\mathbf{-}i\omega t$ time dependence, the imaginary part of passive (lossy) materials is \textbf{positive}. Finally, the so-called permittivity of the rods (and the rods only) can be $z$-anisotropic, $i.e.$ 
		of the form given in Eq.~(\ref{eq:tenseps}).
		\begin{itemize}
			\item[$\bullet$] \onelabentry{\checkboxC}{Enable anisotropy for rods?}  
											 If checked, the permittivity of the rods (and the rods only) 
											 will be $z$-anisotropic with values given below.
											 Checking this will override material rods above.
			\item[$\bullet$] \onelabentry{value}{epsilonr XX re} sets $\RE{\{\varepsilon_{xx}\}}$
			\item[$\bullet$] \onelabentry{value}{epsilonr XX im} sets $\IM{\{\varepsilon_{xx}\}}$
			\item[$\bullet$] \onelabentry{value}{epsilonr YY re} sets $\RE{\{\varepsilon_{yy}\}}$
			\item[$\bullet$] \onelabentry{value}{epsilonr YY im} sets $\IM{\{\varepsilon_{yy}\}}$
			\item[$\bullet$] \onelabentry{value}{epsilonr ZZ re} sets $\RE{\{\varepsilon_{zz}\}}$
			\item[$\bullet$] \onelabentry{value}{epsilonr ZZ im} sets $\IM{\{\varepsilon_{zz}\}}$
			\item[$\bullet$] \onelabentry{value}{epsilonr XY re} sets $\RE{\{\varepsilon_{xy}\}}$
			\item[$\bullet$] \onelabentry{value}{epsilonr XY im} sets $\IM{\{\varepsilon_{xy}\}}$
		\end{itemize}
		Note that so-called $z$-anisotropy means for the relative permittivity tensor that
		$\varepsilon_{xz}=\varepsilon_{yz}=\varepsilon_{zx}=\varepsilon_{zy}=0$ and that 
		$\varepsilon_{yx}=\overline{\varepsilon_{xy}}$.
\subsection{{\ttfamily Incident plane wave}}
	\begin{itemize}
		\item[$\bullet$] \onelabentry{value}{wavelength [nm]}
										 sets the operating freespace wavelength
										 $\lambda_0$ of the incident plane wave.
		\item[$\bullet$] \onelabentry{value}{incident plane wave angle [deg]}
								  		sets the angle of incidence $\theta^i$ of the incident plane wave.
		\item[$\bullet$] \onelabentry{\DOWNarrow\,\,menu}{polarization case}
			 								allows to select the (scalar) $H^\parallel$ or  $E^\parallel$ polarization cases (see Eq.~(\ref{eq:incPW})), or the conical 2.5D case (see Chap.~\ref{part:conical}).
		\item[$\bullet$] \onelabentry{value}{number of post-processed diffraction orders}
										 sets the number of diffraction orders to be post-processed (\emph{e.g.} if set to 2,
										 five Fourier coefficients will be computed corresponding 
										 to diffraction orders -2,-1,0,+1,+2)
	\end{itemize}
\subsection{{\ttfamily Mesh size and PMLs parameters}}
\begin{figure}[H]
		\centering
    \subfloat[\onelabentry{1}{refinement rods [-]}] {\includegraphics[draft=\flagdraft,width=.4 \textwidth]{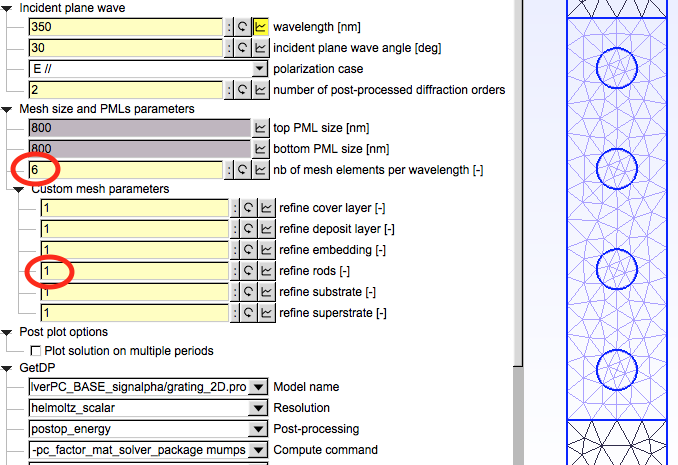}}\quad
    \subfloat[\onelabentry{3}{refinement rods [-]}] {\includegraphics[draft=\flagdraft,width=.4 \textwidth]{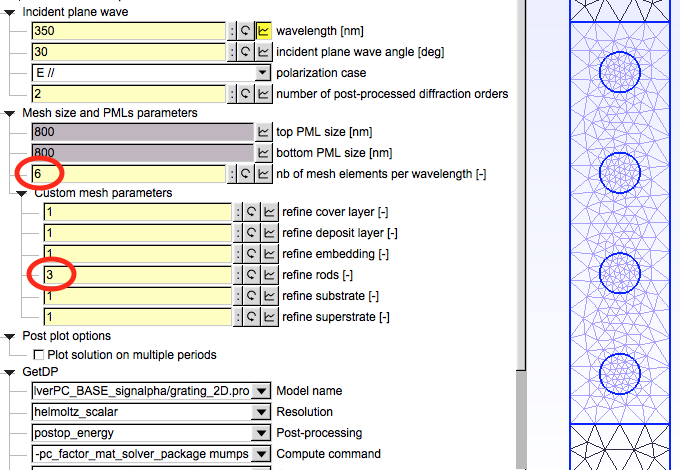}}
		\caption{Mesh refinement options.}\label{fig:meshrefine}
	\end{figure}

	\begin{itemize}
		\item[$\bullet$] \onelabentry{value}{top PML size [nm]}
											allows to set the top PML thickness.
											Typically, it should not be set to a value smaller than $\lambda_0/2$ while a value larger than $3\lambda_0$ is pointless ; $\lambda_0$ is usually a reasonable value with the default PML complex stretch.
		\item[$\bullet$] \onelabentry{value}{bottom PML size [nm]}
											allows to set the top PML thickness.
											Typically, it should not be set to a value smaller than $\lambda_0/2$ while a value larger than $3\lambda_0$ is pointless ; $\lambda_0$ is usually a reasonable value with the default PML complex stretch.
		\item[$\bullet$] \onelabentry{value}{nb of mesh elements per wavelength [-]}
											sets the average number of triangles used to discretize one freespace
											wavelength (mesh refinement).
											Typically, setting it to 30 offers 4 or 5 significant 
											digits over energy related quantities while setting it to 1 leads to very wrong results\dots
		\item[$\bullet$] \stt{Custom Mesh parameters}:
											When dealing with metals and/or very small objects, it is sometimes necessary to
											locally refine the mesh in the affected subdomain, which can 
											be prescribed in this section.
											For instance, Figs.~\ref{fig:meshrefine} shows a local refinement of the rods.
											In Fig.~\ref{fig:meshrefine}(a), the mesh is very coarse, its typical size 
											is $\lambda_0/6$ everywhere. The mesh size within the rods in
											Fig.~\ref{fig:meshrefine}(b) is 3 times smaller ($\lambda_0/(6\times 3)$).
	\end{itemize}

\subsection{{\ttfamily Post plot options}}
\begin{figure}[H]
	\centering
		\includegraphics[draft=\flagdraft,width=.7\textwidth]{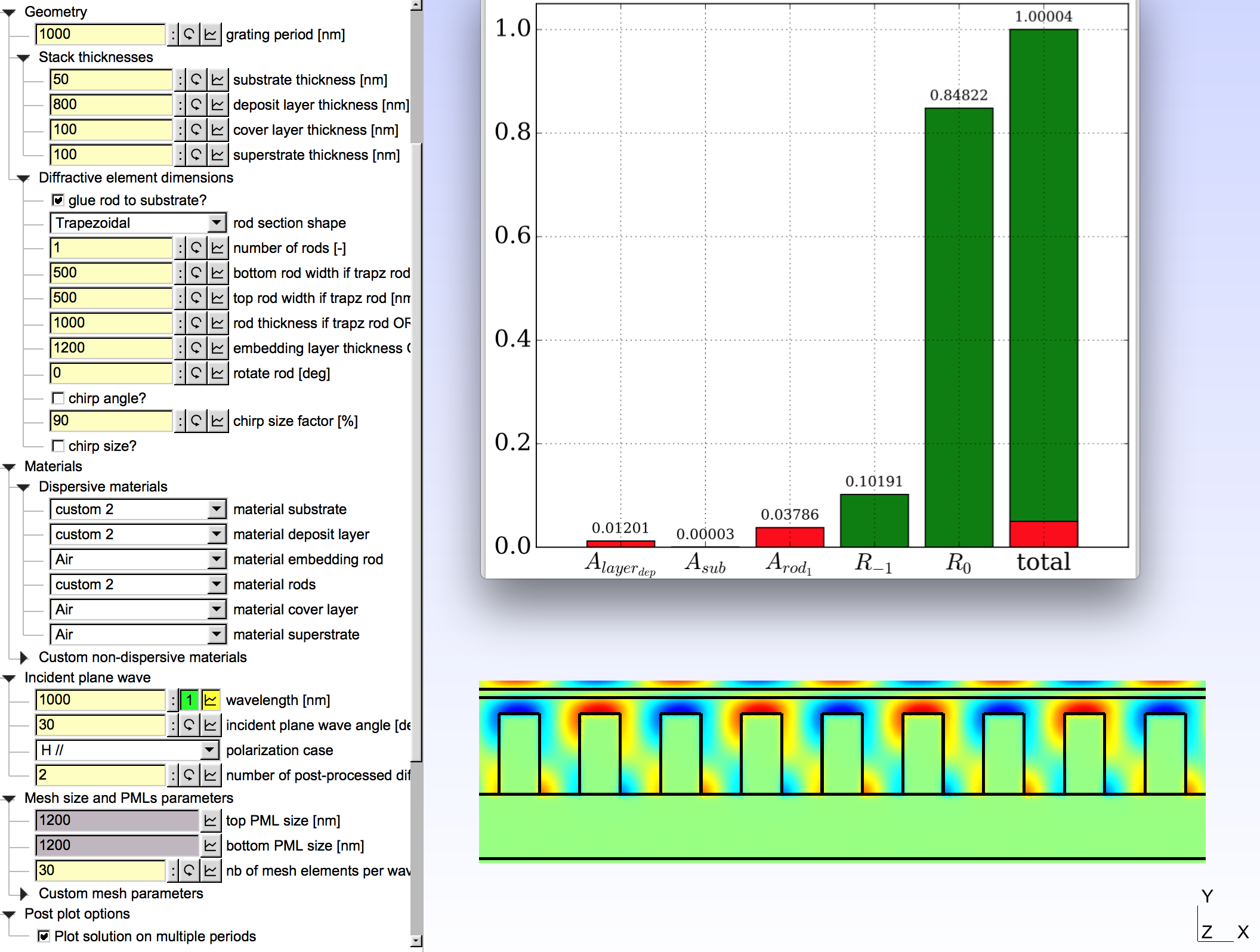}
	\caption{Lamellar grating example. The bar plot shows the output of \stt{grating\_2D\_postplot.py}.}\label{fig:lamellar}
\end{figure}

	\begin{itemize}
		\item[$\bullet$] \onelabentry{\checkboxC}{Plot solution on multiple periods.}
			If checked, the field ($E_z(x,y)$ in $E^\parallel$ polarization case, $H_z(x,y)$ in
			 $H^\parallel$ polarization case)
			is post-processed over 9 grating periods cells, as shown in Fig.~\ref{fig:lamellar}. The field 
			in a neighboring cell is indeed nothing but the field in the reference cell up 
			to a phase shift $e^{\pm i \alpha d}$.
	 \end{itemize}

\section{Energy balance post-processing in {\ttfamily python}}
The provided file \stt{grating\_2D\_postplot.py} gives a possible representation of energy related quantities. If only a single \stt{ONELAB} run was made, it provides bar plot of non-null absorption, reflection and transmission. If a parametric \stt{ONELAB} run was made, \emph{e.g.} a spectrum, it provides a plot of non-null absorption, reflection and transmission.

\section{Examples}
In this section, various example of the literature are retrieved.
\subsection{General recommendations.}
The \stt{ONELAB} model internal files, \stt{grating\_2D.geo} and \stt{grating\_2D.pro}. Both call a configuration file named \stt{grating\_2D\_data.geo} setting all the parameters displayed in the \stt{gmsh} left panel. Thus in order to load directly one of the provided configurations, just rename \stt{grating\_2D\_data\_someconfig.geo} to \stt{grating\_2D\_data.geo} (and \stt{grating\_2D\_data.geo} to \stt{grating\_2D\_data\_old.geo}). Then, open \stt{grating\_2D.pro} with \stt{gmsh}. It is advised to clean the working directory between two different study. To do so, remove at least the output directory \stt{run\_results} and the mesh file \stt{grating\_2d.msh} need to be deleted. Rarely, the Bloch boundary condition fails and \stt{getdp} will complain not finding twin nodes on the two boundaries. Just change the mesh parameter a little and \dots remesh. Finally, if you are not satisfied with the numerical precision, try to refine the mesh and/or increase the size of the PMLs.

\subsection{Lamellar grating example.}
The \stt{LamellarGrating} example (parameter file \stt{grating\_2D\_data\_LamellarGrating.geo}) reproduces some results found in lower half of Table~n$^\circ$2 in \cite{granet1999reformulation}.
At least three significant digits are obtained on the diffraction efficiencies.

\subsection{Anisotropic grating}
\begin{figure}[H]
	\centering
	\includegraphics[draft=\flagdraft,width=.7\textwidth]{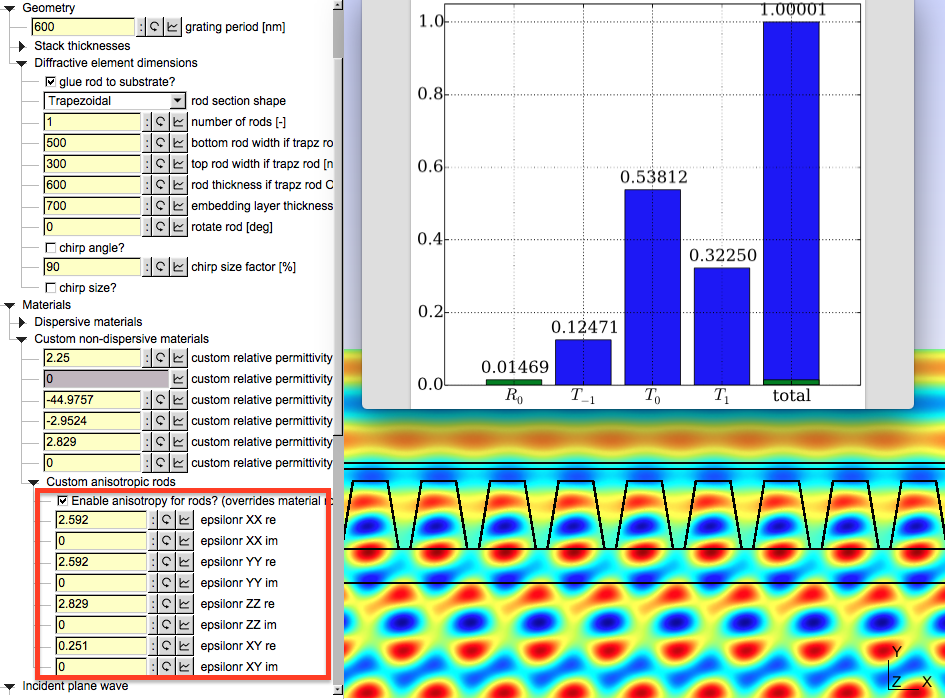}
	\caption{$H^\parallel$ case. See Fig.~5b and 6$^{th}$ line of Tab.~2 in 
    reference \cite{demesy2007thefinite}.}\label{fig:trapzgrating0}
\end{figure}

\begin{figure}[H]
	\centering
	\includegraphics[draft=\flagdraft,width=.7\textwidth]{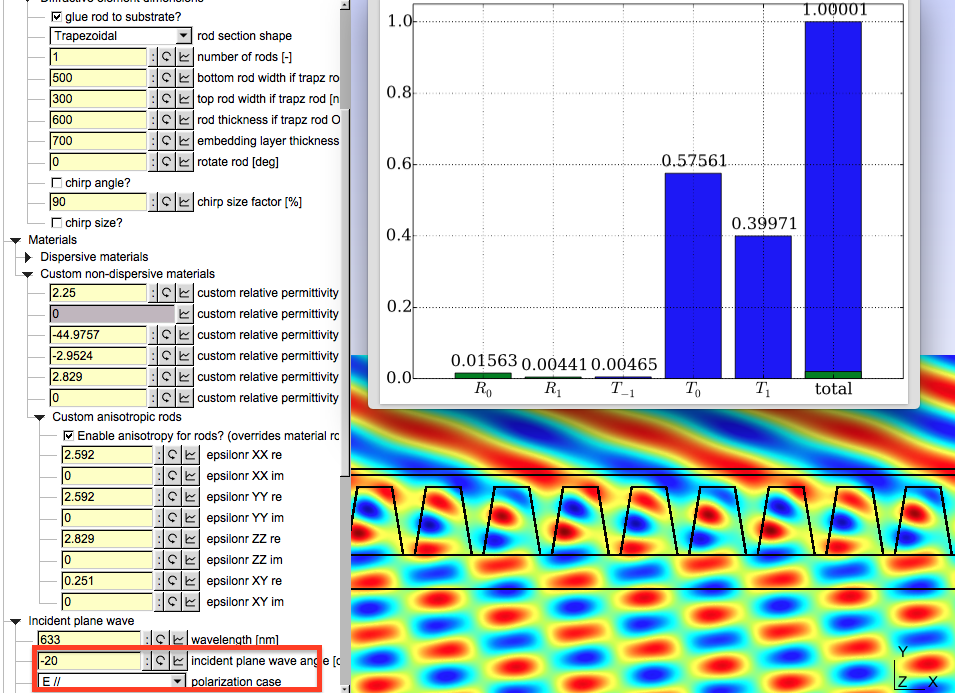}
	\caption{$E^\parallel$ case. See Fig.~5c and third line of Tab.~2 in 
    reference \cite{demesy2007thefinite}.}\label{fig:trapzgrating1}
\end{figure}

The \stt{AnisotropicGrating} example (parameter file \stt{grating\_2D\_data\_AnisotropicGrating.geo}) illustrates the numerical results in \cite{demesy2007thefinite}. Figure~\ref{fig:trapzgrating0} shows the field map $\RE\{E_z\}$ in $V/m$ for an angle of incidence  $\theta^i=-20^\circ$. There is no anisotropic behavior here since $E_z$ only \enquote{sees} $\varepsilon_{zz}$ Figure~\ref{fig:trapzgrating1} shows the field map $\RE\{H_z\}$ in $A/m$ for an angle of incidence $\theta^i=0^\circ$. The lack of symmetry due to the anisotropy of the scatterer is clearly visible here.

\subsection{Photonic crystal slab example.}
This \stt{PhotonicCrystalSlab} example (parameter file \stt{grating\_2D\_data\_PhotonicCrystalSlab.geo}) illustrates some results found in the textbook~\cite{joannopoulos2008molding} (see Fig.~2, page~68). In this example, the band structure of a 2D photonic crystal is given in the two polarization cases. The $E^\parallel$ case features a full photonic bandgap. As a consequence, a sufficiently thick slice of this infinite crystal is expected to exhibit good reflecting properties for an incident plane wave with frequency within the bandgap. The photonic crystal is made of circular rods of diameter $0.4a$ with relative permittivity $\varepsilon_r=8.9$ arranged in a square lattice with lattice constant $a$, with background relative permittivity $\varepsilon_r=1$. The $E^\parallel$ gap is found to be roughly in the normalized frequency $\omega a/2 \pi c$ range $[0.3,0.43]$. In other words, setting in the period $d$ to $150\,$nm should place the bandgap in the wavelength range $[440,660]\,$nm. The gap is total so the reflection on a slab with a few lattices should lead to high reflection for any angle of incidence.

\begin{figure}[H]
	\centering
	\subfloat[Configuration and total field.]
		{\includegraphics[draft=\flagdraft,width=.55\textwidth]{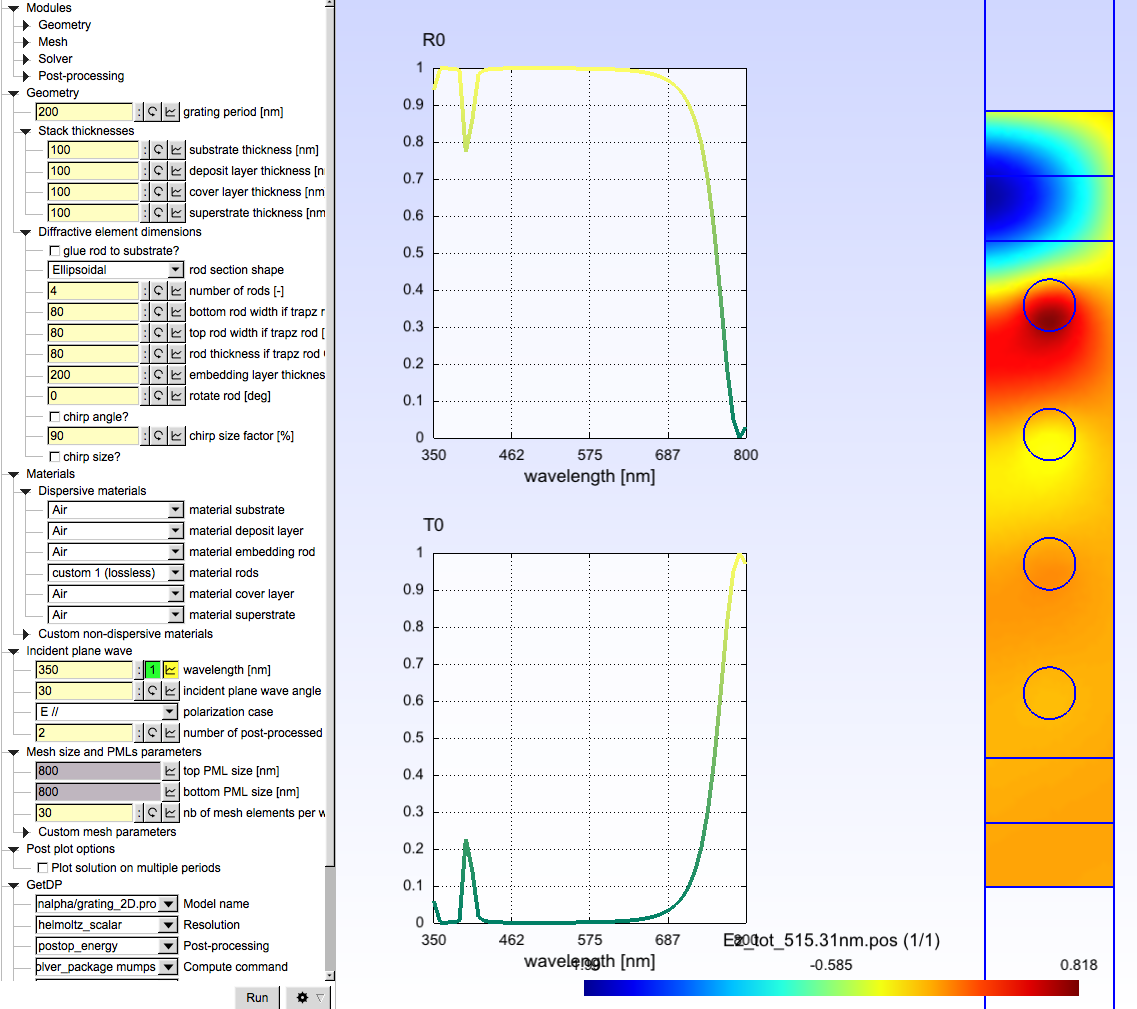}}\quad
	\subfloat[Energy balance output from \stt{grating\_2D\_postplot.py}]
		{\includegraphics[draft=\flagdraft,width=.42\textwidth]{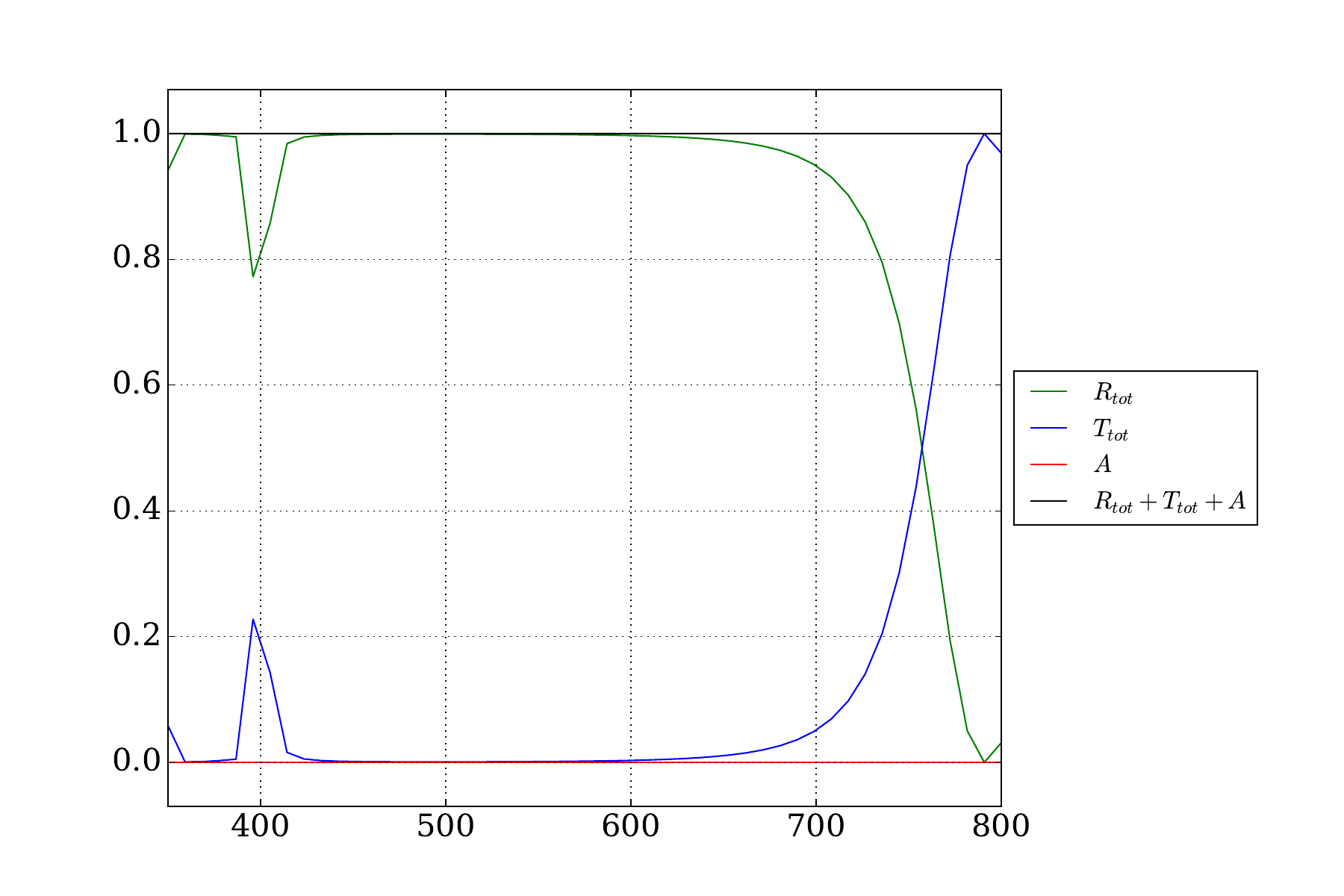}}
	\caption{Photonic crystal slab example.}\label{fig:gap}
\end{figure}

\begin{figure}[H]
	\centering
		\includegraphics[draft=\flagdraft,width=.7\textwidth]{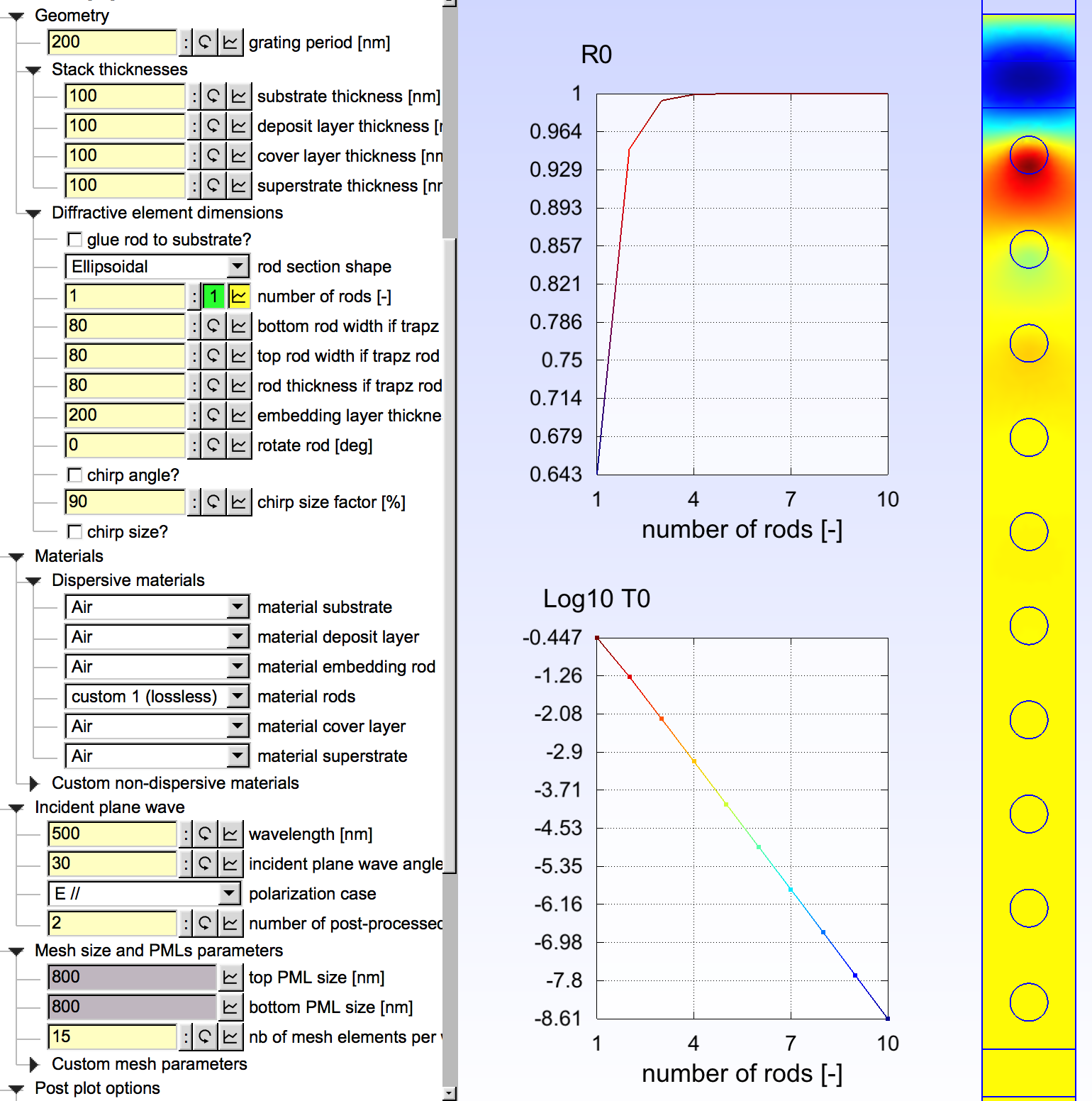}
		\caption{Parametric study as a function of $N_{rods}$.}\label{fig:gapnrods}
\end{figure}

As depicted in Fig.~\ref{fig:gap}(a), for an angle of incidence $\theta^i=30^\circ$, a very high reflection coefficient is obtained for $N_{rods}=5$ only. The python program \stt{grating\_2D\_postplot.py} produces the figure in Fig.~\ref{fig:gap}(b).
Finally, one wonders the slab thickness necessary to witness a high reflectivity (\emph{i.e.}
how many periods in $y$ do we need to see the gap?). A possible parametric study is possible by simply:
\begin{itemize}
  \item unchecking \enquote{looping over}  $\lambda_0$ \onelabbuttons{gray!30}{gray!30}, setting it to $\lambda_0=500\,$nm,
  \item checking \enquote{looping over} $N_{rods}$\onelabbuttons{green}{yellow}, setting looping parameters
	 \button{gray!30}{:} to \mybox{{\sf 1:10:1}},
\end{itemize}
Figure~\ref{fig:gapnrods} shows in log scale the transmission coefficient decaying exponentially with photonic crystal slab thickness. This is expected given the evanescent nature of the field inside photonic crystal slab.

\subsection{Resonant grating}
The \stt{ResonantGrating} example (parameter file \stt{grating\_2D\_data\_ResonantGrating.geo}) illustrates the behavior  of resonant grating that can be used to obtain a very sharp spectral response as detailed in Ref.~\cite{fehrembach2002phenomenological}.
\subsubsection{Spectral response}
\begin{figure}[H]
	\centering
	\includegraphics[draft=\flagdraft,width=.9\textwidth]{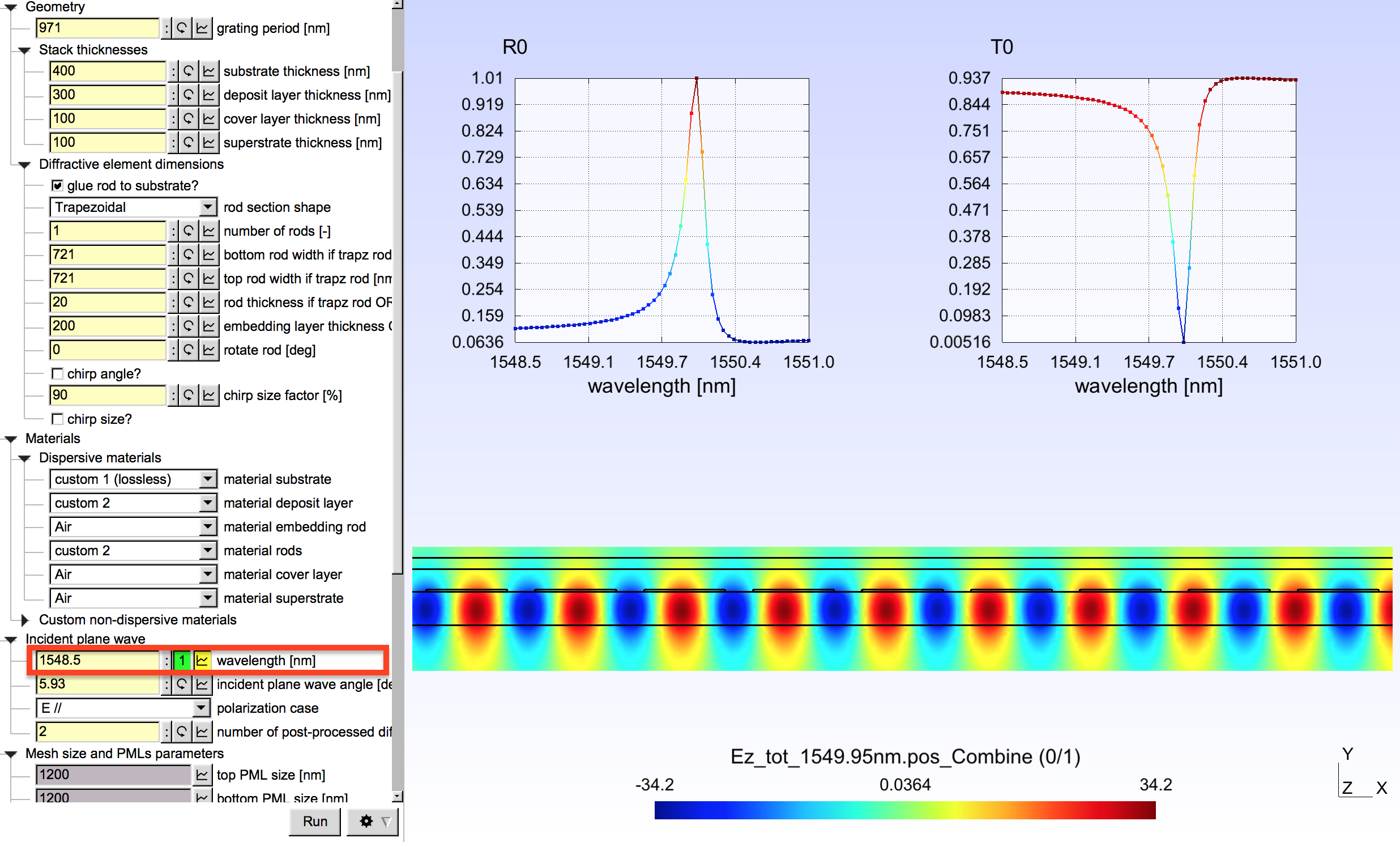}
	\caption{Spectral response.}\label{fig:resonantlambda}
\end{figure}

The spectral response of such a grating is depicted in Fig.~\ref{fig:resonantlambda}.

\subsubsection{Angular response}
\begin{figure}[H]
	\centering
	\includegraphics[draft=\flagdraft,width=.7\textwidth]{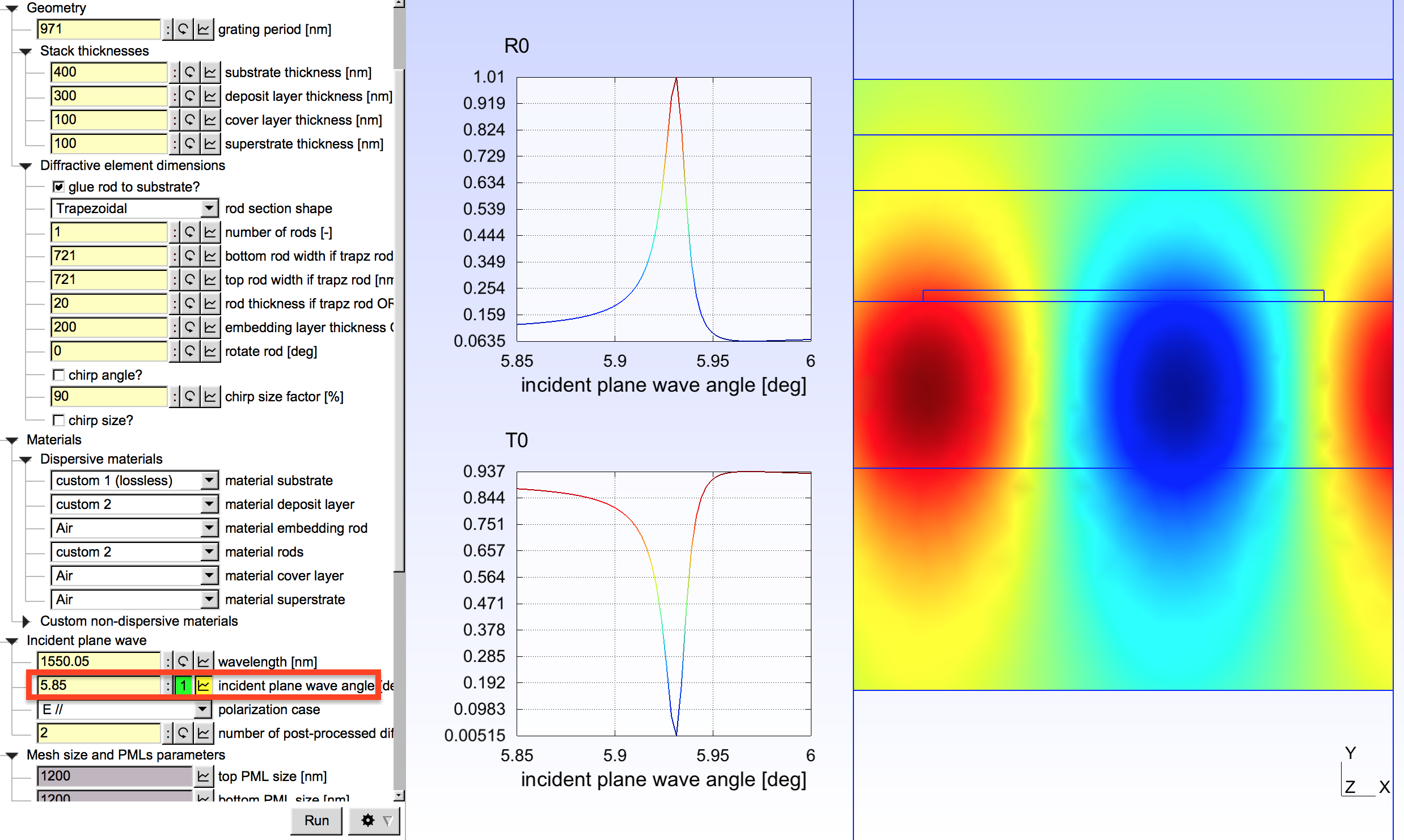}
	\caption{Angular response.}\label{fig:resonantangle}
\end{figure}

From the very same model up to a few preliminary clicks, one can obtain the  angular response of the filter by:
\begin{itemize}
  \item unchecking \enquote{looping over}  $\lambda_0$, setting it to $\lambda_0 = 1550.05\,$nm: \onelabentry{1550.05}{wavelength [nm]}\onelabbuttons{gray!30}{gray!30}
  \item checking \enquote{looping over} $\theta^i$: \onelabentry{6}{incident plane wave angle [deg]}\onelabbuttons{green}{yellow},
	\item setting the looping parameters for $\theta^i$ from $5.85^\circ$ to $6.0^\circ$ by $0.0025^\circ$ steps using
	the button \button{gray!30}{:} and filling \mybox{{\sf 5.85:6:0.0025}}.
\end{itemize}
The angular response of this grating is depicted in Fig.~\ref{fig:resonantangle}.

\subsection{Plasmonic grating}
\begin{figure}[H]
	\centering
	\includegraphics[draft=\flagdraft,width=.9\textwidth]{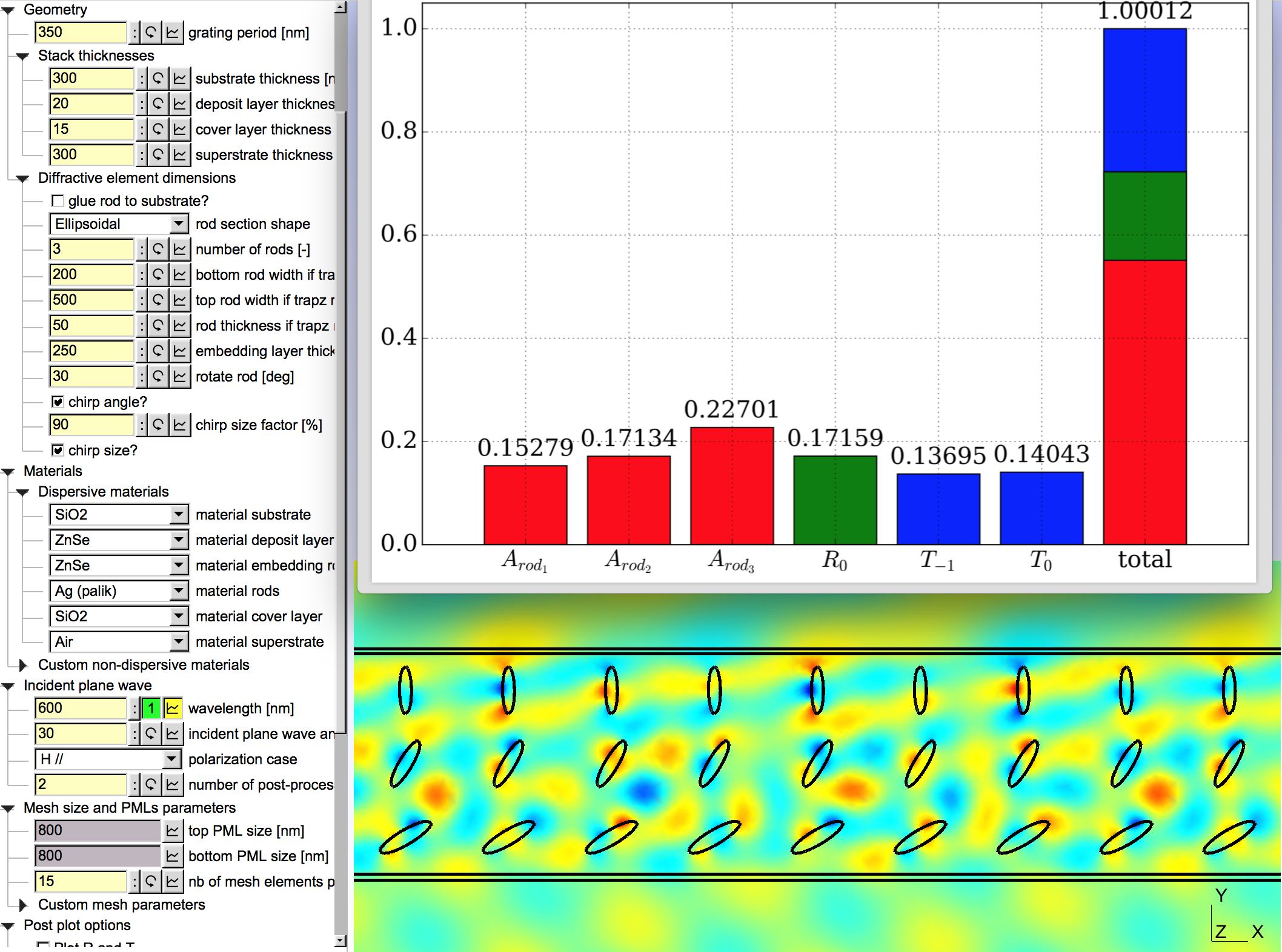}
	\caption{Plasmonic grating.}\label{fig:plasmonicgrating}
\end{figure}

The example in \stt{grating\_2D\_data.geo} has no other purpose than to show that the model handles exotic so-called plasmonic  configurations. The detailed energy balance associated to this weird silver  structure in Fig.~\ref{fig:plasmonicgrating} shows an equilibrated repartition  of the energy occurring inside losses in each rod, reflection and transmission  in both specular and non-specular diffraction efficiencies.

\section{Conclusion}
This model is a general tool for the study of so-called mono-dimensional grating. Various geometries and materials can be handled or easily added. For instance, it can be easily adapted to nano-structured solar cells. The two classical polarization cases, denoted here $E^\parallel$ and $H^\parallel$, are  addressed. The output consists in a full energy balance of the problem computed from  the field maps.

\let\subparagraph\paragraph
\let\paragraph\subsubsection
\let\subsubsection\subsection
\let\subsection\section
\let\section\chapter
\chapter{Crossed gratings : {\tt grating3D.pro}}
\subsection{Problem statement}
\subsubsection{Structure and notations}
We denote by $\bxh$, $\byh$ and $\bzh$ the unit vectors of the axes of an
orthogonal coordinate system $Oxyz$. We only deal with time-harmonic
fields; consequently, electric and magnetic fields are represented
by the complex vector fields $\bE$ and $\bH$, with a time dependance
in $\exp(-i\omega t)$. 
\begin{figure}[t]
	\centering
	\includegraphics[width=.6\textwidth]{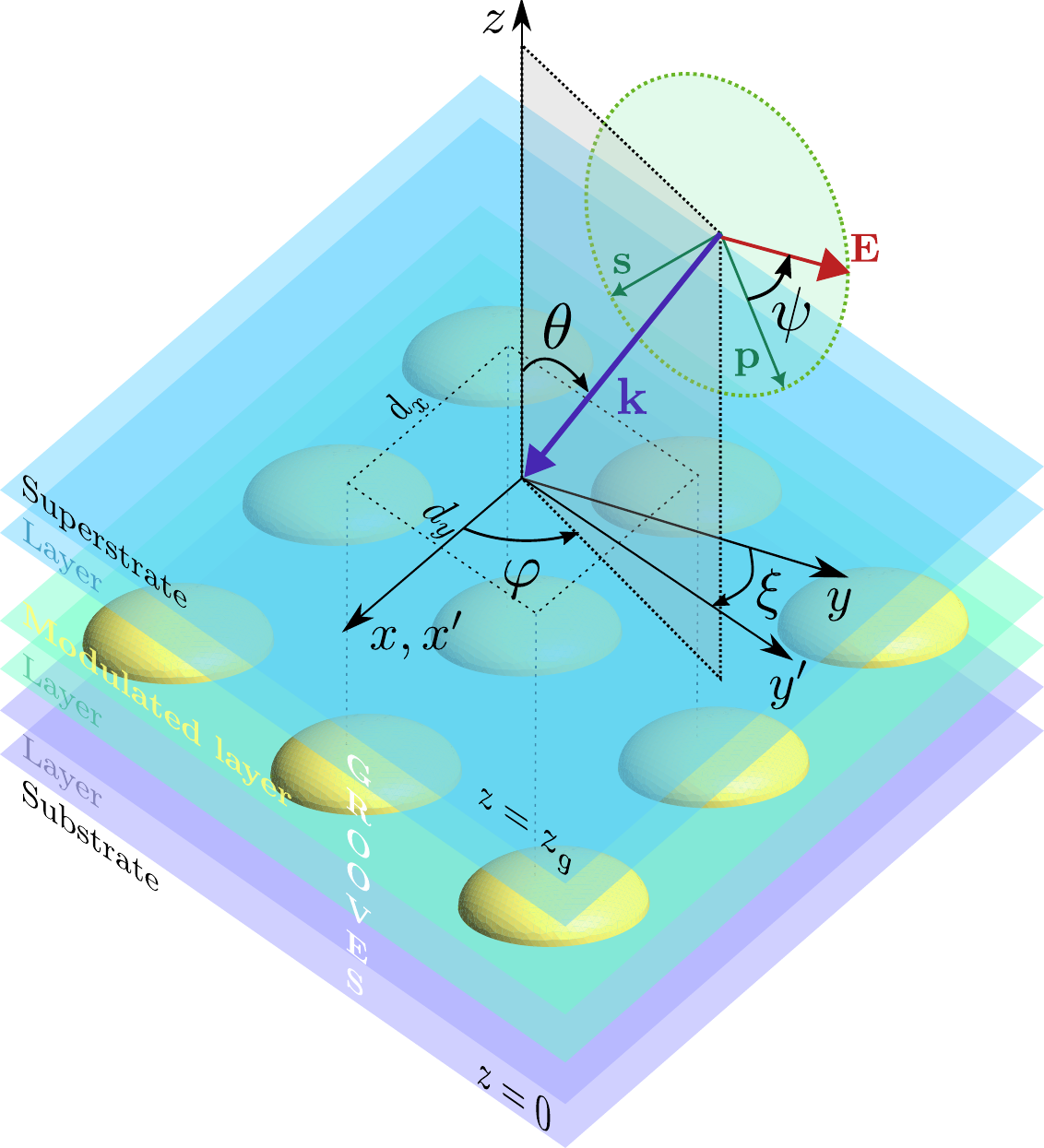}
	\caption{Scheme of the crossed grating and angles convention for the incident plane wave.}
	\label{fig:grating3D}
\end{figure}
Besides, in this section, for the sake of simplicity, the substrate and superstrate are assumed to be isotropic. It is of importance to note that lossy materials can be studied, the relative permittivity and relative permeability being represented by complex valued functions. As detailed in the introduction of this chapter, the annex problem allowing to define a proper scattered field formulation is the diopter one, schematically depicted in Fig.~\ref{fig:locsources}(b). Consequently, the tensor fields $\tensepsr$ and $\tensmur$ fully characterizing the opto-geometric characteristics of the crossed-gratings we are dealing with (see Fig.~\ref{fig:grating3D} where each color represents possibly distinct materials) can be defined by part over the following regions:
\begin{itemize} 
	\item \textit{The  superstrate} ($z>z_g$) is
		supposed to be homogeneous, isotropic and lossless, and therefore characterizedby its relative permittivity $\epsrin{1}$ and its relative permeability $\murin{1}$ and we denote $k_1:=k_0\, \sqrt{\epsrin{1} \murin{1}}$, where $k_0:=\omega/c$,%
	\item \textit{The groove region} ($0<z<z_{g}$), which is possibly heterogeneous and/or anisotropic. The relative permittivity and permeability can vary continuously (gradient index gratings) or discontinuously (step index gratings). It means that the groove region can be constituted of a multilayer stack for instance. This region is thus characterized by the
		tensor fields $\tensepsr^{g}(\bx)$ and $\tensmur^{g}(\bx)$. The
		groove periodicity along the $x$--axis, respectively (resp.) $y'$--axis, is denoted $d_x$, resp. $d_y$.
	\item \textit{The substrate} ($z<0$) is supposed to be homogeneous
		and isotropic and therefore characterized by its relative
		permittivity $\epsrin{2}$ and its relative permeability
		$\murin{2}$ and we denote $k_2:=k_0\,\sqrt{\epsrin{2} \murin{2}}$,%
\end{itemize}
In short, we have defined :
\begin{equation}
	\tensepsr(\bx)= \left \{
	\begin{array}{ll}
	  \epsrin{1}\,\tensid & \mbox{for $z>z_g$}  \\
	  \tensepsr^{g}(\bx)  & \mbox{for $z\in[0,z_g]$} \\
	  \epsrin{2}\,\tensid & \mbox{for $z<0$}
	\end{array}
	\right .
	\mbox{ and }
	\tensmur(\bx)= \left \{
		\begin{array}{ll}
		  \murin{1}\,\tensid & \mbox{for $z>z_g$}  \\
		  \tensmur^{g}(\bx)  & \mbox{for $z\in[0,z_g]$} \\
		  \murin{2}\,\tensid & \mbox{for $z<0$}
		\end{array}
		\right . .
\end{equation}
\subsubsection{Incident plane wave}
The incident field on this structure is denoted:
\begin{equation}\label{eq:def_Einc}
	\bE^{\mathrm{inc}} =\textbf{A}_0^e\,\textrm{exp}(i \,\bk_1\cdot\bx)
\end{equation}
with
\begin{equation}\label{eq:def_kplus}
	\bk_1 = \left[ \begin{array}{l} k_x \\ k_y \\ k_{z,1}
	\end{array} \right] = k_1\,\left[ \begin{array}{l} -\sin\,\theta_0 \,\cos\,\varphi_0 \\-\sin\,\theta_0 \,\sin\,\varphi_0 \\-\cos\,\theta_0
	\end{array} \right].
\end{equation}

\noindent The vector amplitude $\textbf{A}_0^e$ allows to controle the polarization nature of the plane wave (linear defined by the angle $\psi_0$, circular or elliptical) will be specified later on.

\begin{boxgrey}{Remark}
	The plane wave $\bEinc$ is bi-quasi-periodic in the skewed basis. In order to proove it and incidentally to detemine the relevant Bloch phase shifts, it is sufficient to derive $\bEinc(\bx+d_x\bxh)$ and $\bEinc(\bx+d_y\byh'):$
	\begin{itemize}
		\item  $\bEinc(\bx+d_x\bxh)=\textbf{A}_0^e\,\exp(i \,\bk_1\cdot\bx)\,\exp(i\,d_x \, \bk_1\cdot\bxh)=e^{i\,d_x \, k_1^x}\,\bEinc(\bx)$
		\item $\bEinc(\bx+d_y\byh')=\textbf{A}_0^e\,\exp(i \,\bk_1\cdot\bx)\,\exp\left[(i\,d_y \bk_1\cdot(\cos\,\xi\bxh+\sin\,\xi\byh)\right]$\\[2mm]
		\hspace*{2.65cm}$=\exp\left[-i\,d_y\,k_1\sin\,\theta_0\,\sin(\varphi_0+\xi)\right]\,\bEinc(\bx)$
	\end{itemize}
	These phase shifts will be used in the quasi-periodicity constraints. Note that when $\xi$ vanishes, one retrieves the usual phase shift $e^{i\,d_y \, k_1^y}$ along the $y$ direction.
\end{boxgrey}
  
\subsubsection{Problem statement}
We recall here the diffraction problem: Finding the solution of Maxwell equations in harmonic regime \textit{i.e.} the unique solution ($\bE,\bH$) of:
\begin{subequations}\label{eq:Maxwell2}
	\begin{numcases}{}
		\curle\, \bE=i \omega \mu_0\, \tensmur \, \bH \label{eq:MaxwellrotE2}\\
		\curle\, \bH=-i \omega \varepsilon_0\, \tensepsr \, \bE
		\label{eq:MaxwellrotH2}
	\end{numcases}
\end{subequations}
such that the diffracted field satisfies the so-called \textit{Outgoing Waves Condition} (OWC) and where $\bE$ and $\bH$ are quasi-bi-periodic functions with respect to $x$ and $y'$ coordinates. One can choose to compute arbitrarily $\bE$, or $\bH$ since one can be deduced from the other at the cost and associated numerical of a spatial differenciation. Finally, the diffraction problem amounts to looking for the unique solution $\bE$ of the so-called vector Helmholtz propagation equation, deduced from Eqs.~(\ref{eq:MaxwellrotE2},\ref{eq:MaxwellrotH2}):
\begin{equation}\label{eq:helmpb}
	\left \{
	\begin{array}{ll}
		-\curle\left[\tensmur^{-1}\,\curle\,\bE\right] + k_0^2\,\tensepsr\,\bE = \textbf{0} \\[1mm]
		\mbox{with $\bEd:=\bE-\bE_0$ outgoing} \\[1mm]
		\mbox{and $\bE$ quasi-periodic along $(Ox)$ and $(Oy')$}\\[1mm]
	\end{array}
	\right . ,
\end{equation} 
where $\bE_0$ coincides with $\bEinc$ in the superstrate and the groove regions and vanishes in the substrate. 

\subsection{Scattered field formulation}
\noindent The annex problem allowing to define a suitable scattered field formulation can now be introduced. It corresponds to the problem of a simple plane diopter which is the same problem as before if we consider the grove region filled with the same material as the superstrate. We introduce the tensor fields corresponding to this diopter:
\begin{equation}
	\tensepsra(\bx)= \left \{
	\begin{array}{ll}
	  \epsrin{1}\,\tensid & \mbox{for $z>0$}  \\
	  \epsrin{2}\,\tensid & \mbox{for $z<0$}
	\end{array}
	\right .
	\mbox{ and }
	\tensmura(\bx)= \left \{
		\begin{array}{ll}
		  \murin{1}\,\tensid & \mbox{for $z>0$}  \\
		  \murin{2}\,\tensid & \mbox{for $z<0$}
		\end{array}
		\right . .
\end{equation}
We are looking for the unique solution $\bE_1$ of:
\begin{equation}\label{eq:helmanexpb}
	\left \{
	\begin{array}{ll}
		-\curle\left[\tensmura^{-1}\,\curle\,\bE_1\right] + k_0^2\,\tensepsra\,\bE_1 = \textbf{0} \\[2mm]
		\mbox{with $\bE_1^d:=\bE_1-\bE_0$ outgoing}
	\end{array}
	\right . .
\end{equation}
Now the only difficulty is to obtain a closed form for $\bE_1$ in our 3D setting, where it is trivial in 2D since we are talking about the Fresnel coefficients of the diopter. This will be detailed in the last paragraph.

As explained in the introduction of this chapter, the actual unknown of the problem is a field $\bEd{2}$ defined as the difference between $\bE$ and $\bE_1$ and we have: $\bEd{2}=\bE - \bE_1 =\bEd{}-\bEd{1}$. It is important to note that $\bEd{2}$ satisfies the same outgoing condition as $\bEd{}$ \textit{and} $\bEd{1}$. Again, this is a guarantee that no source will be present in the regions with infinite extent in the scattered field formulation. Finally, making use of the definition of $\bEd{2}$ and of the two vector Helmholtz defined above, the propagation equation satified by can be easily obtained:
\begin{equation}\label{eq:gratingstrong}
    \begin{split}
    -\curle\left[\tensmur^{-1}\,\curle\,\bEd{2}\right]+\frac{\om^2}{c^2}\tensepsr\,\bEd{2}= & \frac{\om^2}{c^2}(\tensepsra-\tensepsr)\bE_1 \\&-\curle\left[\left(\tensmura^{-1}-\tensmur^{-1}\right)\,\curle\,\bE_1\right],
    \end{split}
\end{equation}
where the right-hand side is a quasi-bi-periodic source term with support in the whole groove region solely. 

\subsection{The annex problem}
In order to be useful in the context of the FEM, one need to obtain an analytical or semi-analytical expression for the solution of the annex problem. To that extent, we make use of the Fresnel coefficients of course. The assumption of considering isotropic substrate and superstrate is crucial here. It would be possible to consider anisotropic substrates and superstrates, but it would substantially complexify the notions of diffraction efficiencies, outgoing wave conditions and PMLs, incident field and polarization.

The field $\bE_1$ can be relatively easily obtained in the basis formed by the two traditional $({\bph},{\bsh})$ polarization cases, where the following convention is chosen: ${\bsh}:=[\sin\,\varphi_0,-\cos\,\varphi_0,0]^T$ and ${\bph}:={\bsh}\times\bk_1/k_1$, as shown in green color in Fig.~\ref{fig:grating3D}, so that $(\bph,\bsh,\bk_1/k_1)$ form a direct orthonormal basis. But first, one needs to introduce the wave vectors of the transmitted ($\bk_2$) and reflected fields ($\bk_1^r$):

\begin{equation}\label{eq:def_kplus}
	\bk_1^r = \left[ \begin{array}{l} k_x \\ k_y \\ -k_{z,1} \end{array} \right] 
	\mbox{ and } 
	\bk_2 = \left[ \begin{array}{l} k_x \\ k_y \\ k_{z,2} \end{array} \right].
\end{equation}
with $k_{z,2} = -\sqrt{k_0^2\,\epsrin{2}\,\murin{2}-k_x^2-k_y^2}$.

The Fresnel coefficients are classically given by:
\begin{equation}\label{eq:fresnel}
	\left \{
	\begin{array}{lll}
		r_s = \displaystyle\frac{k_{z,1}-k_{z,2}}{k_{z,1}+k_{z,2}} & \mbox{and} & 
		t_s = \displaystyle\frac{2k_{z,1}}{k_{z,1}+k_{z,2}} \\[5mm]
		r_p = \displaystyle\frac{k_{z,1}\,\epsrin{2}-k_{z,2}\,\epsrin{1}} {k_{z,1}\,\epsrin{2}+k_{z,2}\,\epsrin{1}} & \mbox{and} & 
		t_p = \displaystyle\frac{2k_{z,1}\,\epsrin{2}} {k_{z,1}\,\epsrin{2}+k_{z,2}\,\epsrin{1}}
	\end{array}
	\right . .
\end{equation}
\noindent From the Fresnel coefficients, one can readily right the expressions of the fully ${\bsh}$-polarized electric and magnetic fields :
\begin{equation}\label{eq:sfields}
	\left\{
	\begin{array}{l}
		\bE_{\bsh}^i=\;\;\;\,\exp[i\bk_1\cdot\bx]\,{\bsh}  \\[2mm]
		\bE_{\bsh}^r=r_s\,\exp[i\bk_1^r\cdot\bx]\,{\bsh} \\[2mm]
		\bE_{\bsh}^t=t_s\,\exp[i\bk_2\cdot\bx]\,{\bsh}
	\end{array}
	\right. \mbox{ and }
	\left\{
	\begin{array}{l}
		\bH_{\bsh}^i=\;\,1/Z_1\,\exp[i\bk_1\cdot\bx]\,{\bsh}  \\[2mm]
		\bH_{\bsh}^r=r_p\,/Z_1\,\exp[i\bk_1^r\cdot\bx]\,{\bsh} \\[2mm]
		\bH_{\bsh}^t=t_p\,/Z_1\,\exp[i\bk_2\cdot\bx]\,{\bsh}
	\end{array}
	\right.
\end{equation}
with $Z_1= \sqrt{\mu_0/(\varepsilon_0\epsrin{1})}$.

\noindent The purely ${\bph}$-polarized electric field can be deduced from the ${\bsh}$ magnetic field:
\begin{equation}\label{eq:pfields}
	\left\{
	\begin{array}{l}
		\bE_{\bph}^i=-\bk_1\times\bH_{\bsh}^i  /(\om\varepsilon_0\epsrin{1}) \\[2mm]
		\bE_{\bph}^r=-\bk_1^r\times\bH_{\bsh}^r/(\om\varepsilon_0\epsrin{1}) \\[2mm]
		\bE_{\bph}^t=-\bk_2\times\bH_{\bsh}^t  /(\om\varepsilon_0\epsrin{2}) 
	\end{array}
	\right. 
\end{equation}
\noindent Finally, the two elementary electric fields polarized solution to the diopter problem as $\bE_{1{\bsh}}$ and $\bE_{1{\bph}}$ as :
\begin{equation}\label{eq:sfields}
	\bE_{1\{{\bsh},{\bph}\}} = \left\{
	\begin{array}{ll}
		\bE_{\{{\bsh},{\bph}\}}^i+\bE_{\{{\bsh},{\bph}\}}^r & \mbox{ for z>0} \\[2mm]
		\bE_{\{{\bsh},{\bph}\}}^t & \mbox{ for z<0}
	\end{array}
	\right. 
\end{equation}

\begin{boxgrey2}{In the end,}	
one can conveniently define a linearly polarized plane wave or a circular (left of right) on the bases formed by the $(\bE_{1{\bsh}},\bE_{1{\bph}})$ fields :
\begin{itemize}
	\item linear with amplitude $A_e$ with angle $\psi_0$ with respect to the plane of incidence (see Fig.~\ref{fig:grating3D}) : 
	\begin{equation}
		\bE_1=A_e\left(\cos\,\psi_0\,\bE_{1{\bph}}-\sin\,\psi_0\,\bE_{1{\bsh}}\right)
	\end{equation}
	\item circular right or left with amplitude $A_e$ : 
	\begin{equation}
		\bE_1=\frac{A_e}{\sqrt{2}}\left(\bE_{1{\bph}}\pm i\,\bE_{1{\bsh}}\right)
	\end{equation}	
\end{itemize}
\end{boxgrey2}

Note that this choice is heavily dictated by the fact that the FEM software GetDP nicely handles the notion of vector fields defined by part and the cross-product between them.

\subsection{Variational formulation}\label{part:form_faible_3D}
The variational form residue is obtained by multiplying scalarly Eq.~(\ref{eq:gratingstrong}) by weighted vectors $\bW$ chosen among the ensemble of quasi-periodic square integrable fields with square integrable $\curl$, denoted $H^1\left(\Omega,\curle,(d_x\bxh,d_y\hat{\mathbf{y}}'),\bk\right)$. The variational diffraction problem reads as follows.
\begin{boxgrey2}{$\mbox{Find }\bEd{2}\mbox{ such that }\forall\,\bW\in H^1\left(\Omega,\curle,(d_x\bxh,d_y\hat{\mathbf{y}}'),\bk\right)$}
	\begin{equation}\label{eq:gratingweak}
		\begin{split}
			-&        \GalerkinV{\tensmur^{-1}\,\curle\,\bEd{2}}{\curle\,\bW}{\Omega}
			   +k_0^2 \GalerkinV{\tensepsr\,\bEd{2}}{\bW}{\Omega} \\
			   -&k_0^2 \GalerkinV{(\tensepsra-\tensepsr)\,\bE_1}{\bW}{\Omega_g}\\
			+&\GalerkinV{(\tensmura^{-1}-\tensmur^{-1})\,\curle\,\bE_1}{\curle\,\bW}{\Omega_g} \\
			+&\GalerkinS {\left(\n_{\mathrm{ext}} \times \left((\tensmura^{-1}-\tensmur^{-1})\,\curle\,\bE_1\right)\right)}{\bW} {\partial \Omega_g}\\
			-&\GalerkinS{\left(\n_{\mathrm{ext}} \times (\tensmur^{-1}\,\curle\,\bEd{2})\right)}{\bW}{\Gamma_{\mathrm{PML}}^+\cup\Gamma_{\mathrm{PML}}^-}\\
		   =&0\,.
		\end{split}
	\end{equation}		
\end{boxgrey2}
The last three terms are usually null at optical frequencies: The two terms involving a contrast of permeability are usually null for amagnetic problems, while the very last term arising from the integration by part of the $\curle$ affects the behavior of the field at the PML endings. There are two obvious choices regarding this boundary term: The first one is to force it to zero (homogeneous Neumann condition, perfect magnetic conductor) by simply forgetting it from the formulation, the second one to assume that the tangential components of the field are null at the PML endings by choosing test function null at the PML endings (Dirichlet condition, perfect electric conductor). The first option should be considered if possible because knowing the values of the field at PML endings allows to assess the efficiency of the PML implemented. The advantage of the second option is that it leads to smaller discrete systems since there are no unknowns at the PML endings.  

From a discrete point of view, the first term in Eq.~\ref{eq:gratingweak} leads to the so-called stiffness matrix, the second term to the mass matrix while the third term leads to the load vector.

\subsection{Energy balance: Diffraction efficiencies and losses}\label{part:efficiencies}
In order to define the diffraction efficiencies, one needs to introduce the components of the wavevectors of the corresponding plane waves :

\begin{equation}\label{eq:alphanm}
\left\{
	\begin{array}{l}
		\alpha_{m,n} = -k_x+\dfrac{2\pi}{d_x}m \\[4mm]
		\beta_{m,n}  = -k_y+\dfrac{1}{\cos\,\xi}\dfrac{2\pi}{d_y}n-\tan\,\xi\,\dfrac{2\pi}{d_x}m\\[4mm]
		\gamma^r_{m,n} = \sqrt{k_0^2\,\epsrin{1} - \alpha_{m,n}^2 - \beta_{m,n}^2}\\[4mm]
		\gamma^t_{m,n} = \sqrt{k_0^2\,\epsrin{2} - \alpha_{m,n}^2 - \beta_{m,n}^2}
  
	\end{array}
	\right. 
\end{equation}

\noindent The classical Rayleigh expansion provides the complex amplitude of each diffraction order:  
\begin{equation}\label{eq:coefffourier}
\left\{
	\begin{array}{l}
		e_{m,n}^{r,u} = \dfrac{1}{d_x d_y \cos\,\xi}\dint_{\Gamma^+}\exp\left[i(\alpha_{m,n}x+\beta_{m,n}y)\right]\,\bEd{}\cdot\hat{\mathbf{u}}\,\d S\\[4mm]
		e_{m,n}^{t,u} = \dfrac{1}{d_x d_y \cos\,\xi}\dint_{\Gamma^-}\exp\left[i(\alpha_{m,n}x+\beta_{m,n}y)\right]\,\bE\cdot\hat{\mathbf{u}}  \,\d S\\[4mm]
	\end{array}
	\right. ,
\end{equation}
with $u$ spans $\{x,y,z\}$ and $\hat{\mathbf{u}}$ spans $\{\bxh,\byh,\bzh\}$ and where $\Gamma^+$ (resp. $\Gamma^-$) is a cut of the periodic cell at a constant altitude $z$ with $z>z_g$ (resp. $z<0$). Finally, one can deduce the diffraction efficiencies from the transverse components of the field $\bE\cdot\bxh$ and $\bE\cdot\byh$ only : 
\begin{boxgrey}{Diffraction efficiencies $\parallel$}
\begin{equation}\label{eq:RT1}
	\left\{
		\begin{array}{l}
			R^\parallel_{m,n}=\dfrac{1}{\gamma^r_{m,n}\,k_{z,1}}
			\left[
				({\gamma^r_{m,n}}^2+{\alpha_{m,n}}^2)\,|e_{m,n}^{r,x}|^2
			   +({\gamma^r_{m,n}}^2+{\beta_{m,n}}^2) \,|e_{m,n}^{r,y}|^2\right.\\
			\left.\hspace*{4cm}+2\,\alpha_{m,n}\beta_{m,n}\,\RE\{e_{m,n}^{r,x}\,\overline{e_{m,n}^{r,y}}\}
			\right]\\
			T^\parallel_{m,n}=\dfrac{1}{\gamma^t_{m,n}\,k_{z,1}}
			\left[
				({\gamma^t_{m,n}}^2+\alpha_{m,n}^2)\,|e_{m,n}^{t,x}|^2
			   +({\gamma^t_{m,n}}^2+\beta_{m,n}^2)\,|e_{m,n}^{t,y}|^2\right.\\
			\left.\hspace*{4cm}+2\,\alpha_{m,n}\beta_{m,n}\,\RE\{e_{m,n}^{t,x}\,\overline{e_{m,n}^{t,y}}\}
			\right]
		\end{array}
	\right. ,
\end{equation}
\end{boxgrey}

\noindent This last expression of the diffraction efficiencies is handy since it does not require to explicitly compute the Rayleigh coefficients $e_{m,n}^{r,z}$ and $e_{m,n}^{t,z}$ involving the normal component of the field on $\Gamma^\pm$. Indeed, with vector elements defined on edges and faces of tetrahedrons, the (discontinuous) normal component to a surface is not readily available. It can however be retrieved through the use of a Lagrange multiplier (and the \texttt{Trace} operator defined in GetDP). Once determined, another formula for the diffraction efficiencies can be used :  

\begin{boxgrey}{Diffraction efficiencies $\perp$}
\begin{equation}\label{eq:RT2}
	\left\{
		\begin{array}{l}
			R^\perp_{m,n}=\dfrac{1}{\gamma^r_{m,n}\,k_{z,1}}
			\left[
				\alpha_{m,n}^2    \,|e_{m,n}^{r,x}|^2
			   +\beta_{m,n}^2     \,|e_{m,n}^{r,y}|^2
			   +{\gamma^t_{m,n}}^2\,|e_{m,n}^{r,z}|^2\right]\\
			T^\perp_{m,n}=\dfrac{1}{\gamma^t_{m,n}\,k_{z,1}}
			\left[
				\alpha_{m,n}^2     \,|e_{m,n}^{t,x}|^2
				+\beta_{m,n}^2     \,|e_{m,n}^{t,y}|^2
				+{\gamma^t_{m,n}}^2\,|e_{m,n}^{t,z}|^2\right]\\
		\end{array}
	\right. ,
\end{equation}
\end{boxgrey}

As for the Joule losses whithin an (isotropic) groove region of relative permittivity $\epsrin{g}(\bx)$, they can be retrived by computing the following ratio:
\begin{boxgrey}{Joule Losses}
\begin{equation}\label{eq:losses}
    Q = \frac{\displaystyle \int_{\Omega_g} \frac{1}{2}\,\omega\,\varepsilon_0\,\IM(\epsrin{g})\,|\bE|^2\,\mathrm{d}\Omega}
             {\displaystyle \int_{\Gamma^+} \frac{1}{2}\RE\{\bEinc\times\overline{\bHinc}\}\cdot -\bzh\,\mathrm{d}\Gamma}
    \,.
\end{equation}
\end{boxgrey}

The numerator in Eq.~(\ref{eq:losses}) clarifies losses in watts by bi-period of the considered crossed-grating and are computed by integrating the Joule effect losses density over the volume $V$ of the lossy element. The denominator normalizes these losses to the incident power, \textit{i.e.} the time-averaged incident Poynting vector flux across one bi-period. Since $\bEinc$ is nothing but a plane wave, this last term is equal to $A_e^2/2\,\sqrt{\epsrin{1}\,\varepsilon_0/\mu_0}\,d_x\,d_y\,\cos\,\xi\cos\,\theta_0$.

\subsection{Total field formulation}\label{sec:formtotalfield}
It should be stated that for gratings, it is relatively easy to implement a total field formulation of the problem using a virtual antenna on $\Gamma^+$. The induced current to impose on the surface is equal $2\bzh\times\bH_1$ \cite{jin2015finite,stratton2007electromagnetic}, which is handled through a Robin condition on $\Gamma^+$ :
\begin{equation}\label{eq:gratingweaktotal}
    \begin{split}
		-&\GalerkinV{\tensmur^{-1}\,\curle\,\bE}{\curle\,\bW}{\Omega}
		       +k_0^2 \GalerkinV{\tensepsr\,\bE}{\bW}{\Omega} \\
		-&2 i\omega\mu_0\GalerkinS{\left(\bzh \times \bH_1\right)}{\bW}{\Gamma^+}\\
	   =&0\,,
	\end{split}
\end{equation}
for an amagnetic problem with homogeneous Neumann conditions at PMLs endings, which allows to disregard the last three terms of the formulation in Eq.~\ref{eq:gratingweak}. The assembly time of the total field formulation is slightly shorter than the scattered field one due to the presence of surface source term instead of a volume one. However, the implementation remains of the same level of difficulty as in the scattered field one since the annex problem still has to be solved in order to compute the annex magnetic field $\bH_1$. 

\subsection{Convergence}\label{part:gratingconvergence}
\begin{figure}[h]
    \includegraphics[draft=\flagdraft,width=\textwidth]{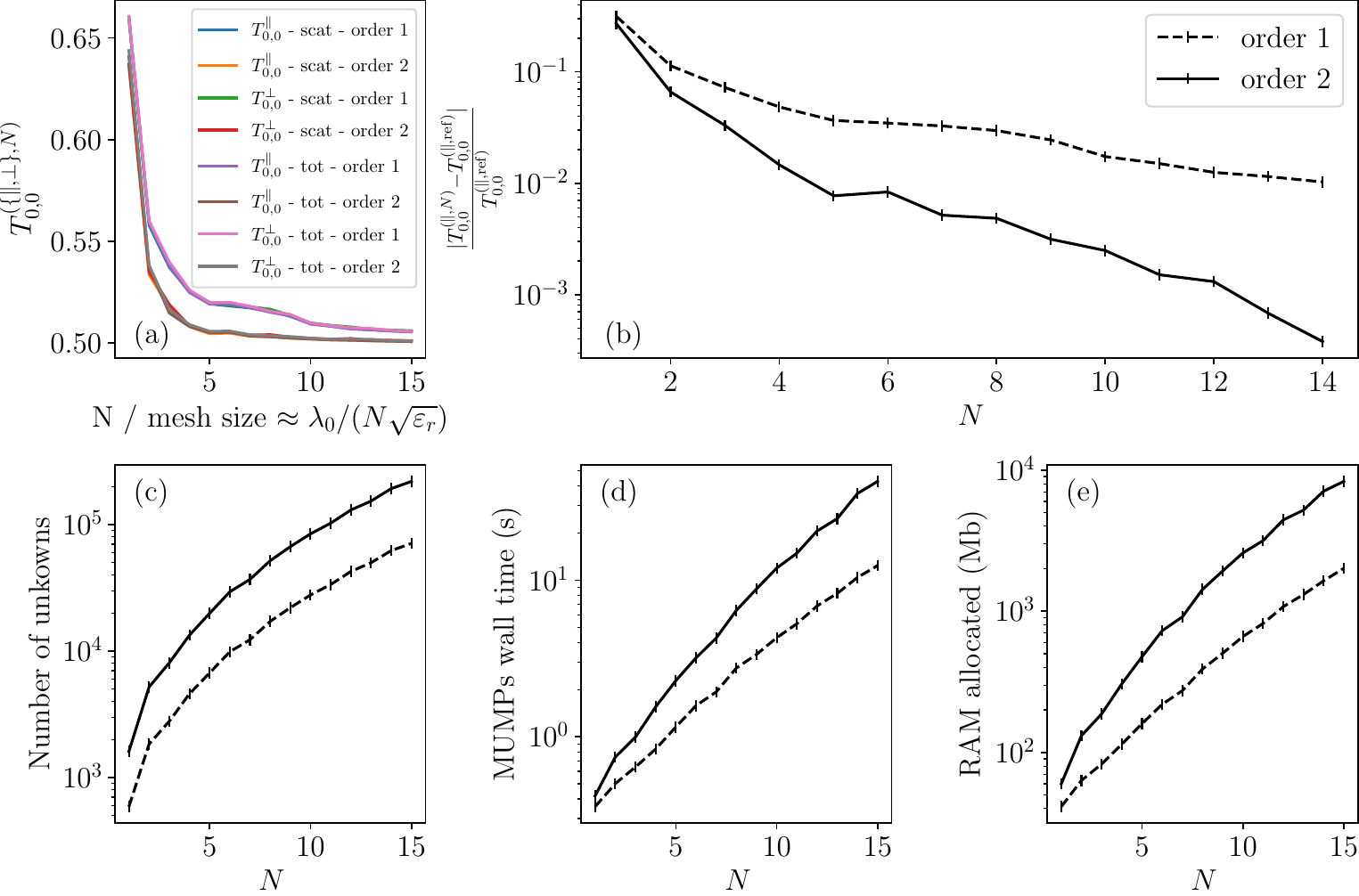}
    \caption{(a) Convergence of $T^\parallel_{0,0}$ (Eq.~\ref{eq:RT1}) and $T^\perp_{0,0}$ (Eq.~\ref{eq:RT2}) as a function of the formulation type (scattered or total), interpolation order and mesh refinement parametrized by $N$, the number of mesh elements per wavelength in a given material. (b) Convergence rate as a function of the mesh refinement and the interpolation order. The reference value $T^{(\parallel,\mathrm{ref})}_{0,0}$ is the one obtained with second order and $N=15$. (c) Number of unknowns as a function of the mesh refinement.  (d) Direct solver wall time in seconds. (e) Allocated RAM.}
    \label{fig:convgrating}
\end{figure}

The convergence of the energy related quantities with respect to the mesh refinement and finite element order can easily be checked using the \texttt{grating3D.pro} model. Running \bashline{gmsh grating3D.pro -setstring test\_case convergence} from the command line allows to retrieve all the results presented in Fig.~\ref{fig:convgrating}. This test case loops over the mesh refinement parametrized by $N$, the approximate number of tetrahedra per wavelength in a given material (\textit{i.e.} the mesh size is set to $\lambda_0/(N\sqrt{\varepsilon_r})$), and the interpolation order.

The blue and orange curves in Fig.~\ref{fig:convgrating}(a) show the specular transmitted efficiency $T^\parallel_{0,0}$ (see Eq.~\ref{eq:RT1}) as a function of $N$. These curves are superimposed with the green and blue ones representing $T^\perp_{0,0}$ (see Eq.~\ref{eq:RT2}). The purple, brown, pink and grey curves represent the same quantities as described just before, when computed using the total field formulation described in Sec.~\ref{sec:formtotalfield}. The convergence rate is shown in Fig.~\ref{fig:convgrating}(b). The corresponding number of unknowns, direct solver runtime and allocated RAM are indicated in Figs.~\ref{fig:convgrating}(c-e). 

\let\subsection\chapter
\subsection{The conical 2.5D case}\label{part:conical}
\begin{figure}[h!]
	\centering
	\includegraphics[draft=\flagdraft,width=.6\textwidth]{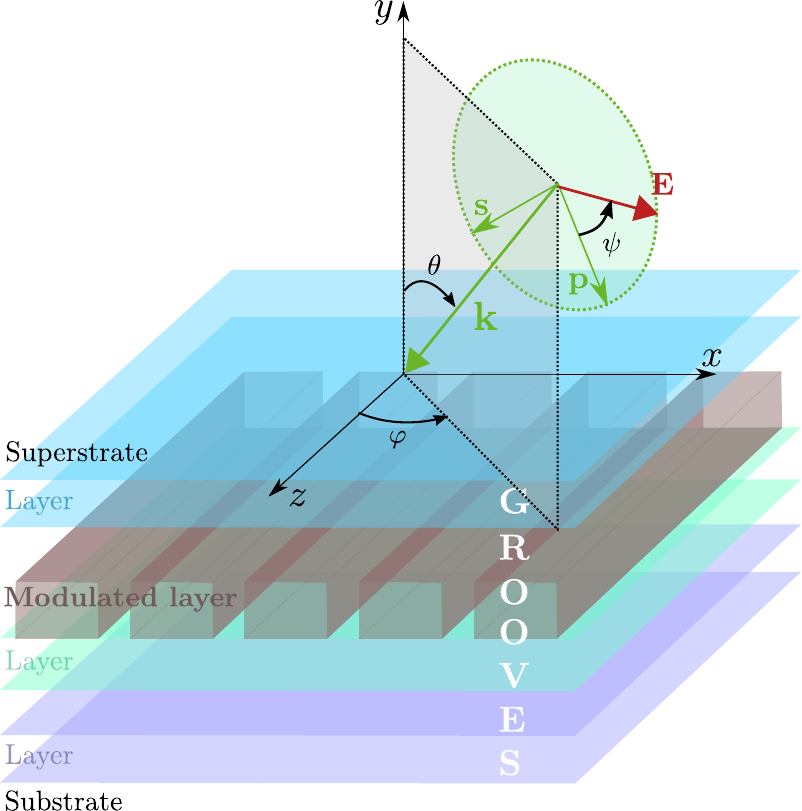}
	\caption{Convention and notation for the conical case.}
	\label{fig:conical}
\end{figure}

The conventions and notations adopted for the conical incidence case, or 2.5D case, are shown in Fig.~\ref{fig:conical} where a 3D plane wave is incident on a 2D geometry. Note that the coordinate system has changed ($x\rightarrow z$, $y\rightarrow x$, $z\rightarrow y$) from the 3D case in order to match the more usual convention where $(Oz)$ is the geometrical axis of invariance. The conical case can be tackled using a mixed formulation \cite{renversez2012foundations}, where the possibly discontinuous transverse components of the unknown field, denoted by $\bE_{2,t}^d:=E_{2,x}^d(x,y)\bxh+E_{2,y}^d(x,y)\byh$, are discretized with edge elements and the continuous longitudinal component denoted $E_{2,z}^d(x,y)$ by nodal elements.

The following handy transverse operators $\gradt$ and $\curlet$ are introduced:
\begin{equation*}
	\gradt\,f=\partial_x f \,\bxh + \partial_y f \,\byh
	\mbox{ and }
	\curlet\,\bF = (\partial_y F_x - \partial_x F_y)\,\bzh .
\end{equation*}

Restricting the generality of the relative permittivity (and permeability) tensors in the following way:
\begin{equation}\label{eq:zaniso}
	\tensepsr = 
	\begin{bmatrix}
		\epsrin{xx} & \overline{\epsrin{a}} & 0\\
		\epsrin{a} & \epsrin{yy} & 0\\
		0 & 0 & \epsrin{zz}
	\end{bmatrix}
	=
	\underbrace{
		\begin{bmatrix}
			\epsrin{xx} & \overline{\epsrin{a}} & 0\\
			\epsrin{a} & \epsrin{yy} & 0\\
			0 & 0 & 1
		\end{bmatrix}
	}_{\Large\tensepsrtt}
	\begin{bmatrix}
		1 & 0 & 0\\
		0 & 1 & 0\\
		0 & 0 & \epsrin{zz}
	\end{bmatrix}
\end{equation}
allows to convieniently decouple the tranverse and longitudinal behavior of the field. Indeed, we are looking for a solution of the scattering problem under the form of the following ansatz $\bEd{2} = \left[\bE_{2,t}^d(x,y) + E_{2,z}(x,y)\,\bzh\right]e^{ik_z z}$, where $k_z$ is the longitudinal componenent of the incident wavevector. Thus, the decoupling mentionned above writes :
\begin{equation*}
	\tensmur^{-1}\curle\,\bEd{2} = \left[\murin{zz}^{-1}\curlet\,\bE_{2,t}^d + \tensmurtt^{-1}
	 (\gradt\,E^d_{2,z}-ik_z\bE_{2,t}^d) \times\bzh\right]e^{ik_z z}
\end{equation*}

Finally, one can obtain the variational formulation for the conical case : 
\begin{equation}\label{eq:conicalweak}
	\setlength\arraycolsep{1pt}
	\begin{array}{ll}
		 & \GalerkinV{ \murin{zz}^{-1}\,\curlet\,\bE_{2,t}^d} {\curlet\,\bW}{\Omega}\\[3mm]
		+&\GalerkinV{ \left(\tensmurtt^{-1}\left(\bzh\times\gradt\,  E_{2,z}^d\right)\right)} {\bzh\times\gradt\,  w}{\Omega} \\[3mm]

		+i k_z&\GalerkinV{\left(\tensmurtt^{-1}\left(\bzh\times\gradt\,E_{2,z}^d\right)\right)} {\bzh\times\bW}{\Omega} \\[3mm]
		-i k_z&\GalerkinV{\left(\tensmurtt^{-1}\left(\bzh\times\bE_{2,t}^d      \right)\right)} {\bzh\times\gradt\,w}{\Omega} \\[3mm]

		+k_z^2&\GalerkinV{ \left(\tensmurtt^{-1}\left(\bzh\times\bE_{2,t}^d\right)\right)} {\bzh\times\bW}{\Omega}\\[3mm]

		 -k_0^2&\GalerkinV{\tensepsrtt\,\bE_{2,t}^d} {\bW}{\Omega}
		 -k_0^2\GalerkinV{\epsrin{zz}\,  E_{2,z}^d} {  w}{\Omega}+\mbox{B.T.}\\[3mm]

		 +k_0^2&\GalerkinV{(\tensepsratt-\tensepsrtt)\,\bE_{1,t}}  {\bW}{{\Omega_g}}
		 +k_0^2\GalerkinV{(\epsrin{a,zz}-\epsrin{zz})\,  E_{1,z}} {  w}{{\Omega_g}} =0, 
	\end{array}
\end{equation}
where B.T. is the boundary term arising from the integration by part of the $\curle$ operator.  
\begin{figure}[h!]
	\includegraphics[draft=\flagdraft,width=.8\textwidth]{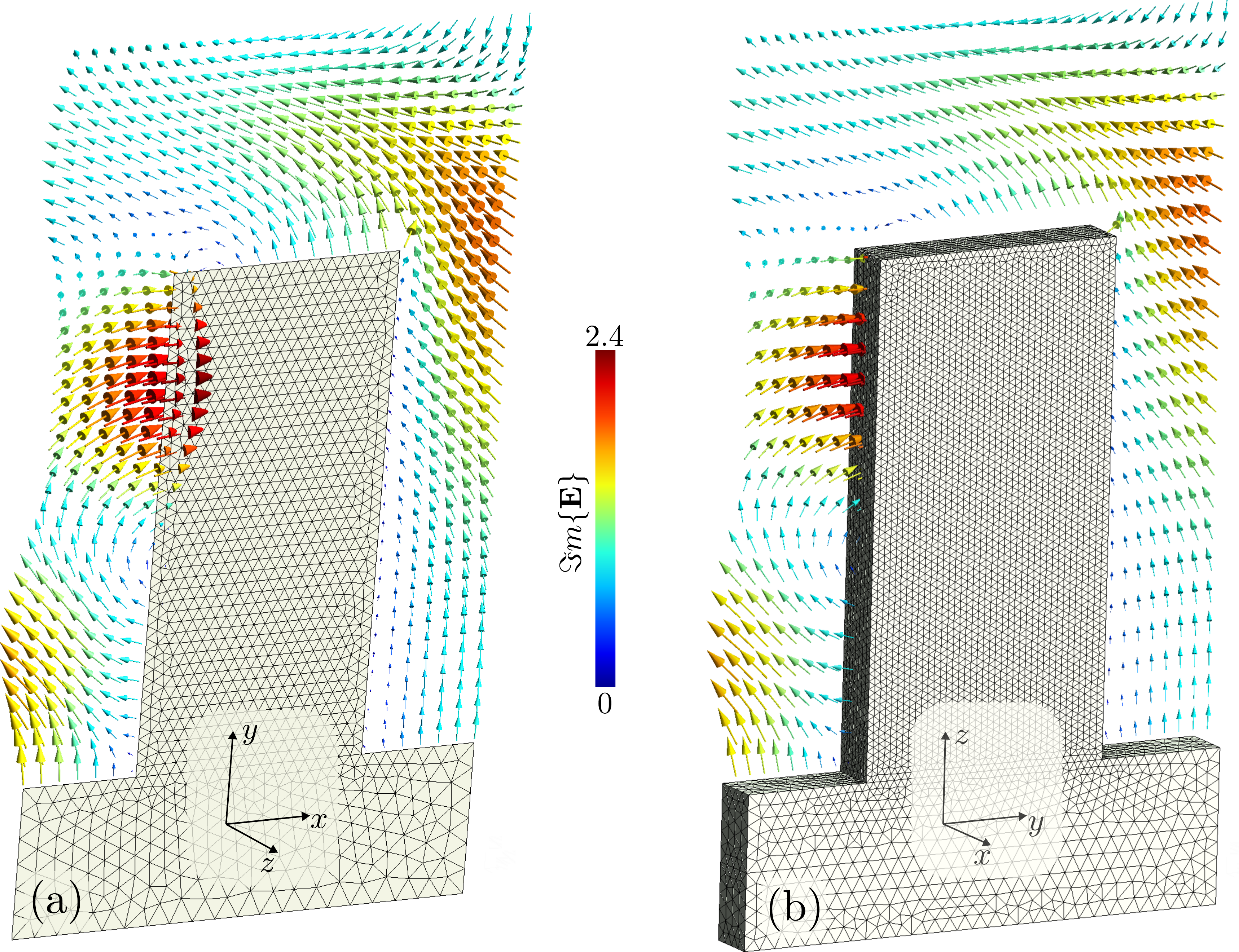}
	\caption{Total electric field obtained with the conical model (a) and with the 3D model (b).}
	\label{fig:compare3Dconical}
\end{figure}
As a validation, the same conical configuration is computed using the full 3D formulation (Fig.~\ref{fig:compare3Dconical}(b)) and with the 2.5D conical formulation (Fig.~\ref{fig:compare3Dconical}(b)). The grating is made of silver on a silver substrate and the plane wave angles are in both cases $\theta_0=\SI{30}{\degree}$, $\varphi_0=\SI{30}{\degree}$ and $\psi_0=\SI{10}{\degree}$. All the parameters can be found in the \texttt{ONELAB} template models \texttt{grating2D.pro} and \texttt{grating3D.pro}. The 3D case can be reproduced by command line : \bashline{gmsh grating3D.pro -setstring test\_case retrieve\_2D\_lamellar} or by opening \texttt{grating3D.pro} in \texttt{Gmsh} and selecting \texttt{retrieve\_2D\_lamellar} in the \texttt{Geometry} drop-down menu. The 2D conical case can be reproduced by command line : \bashline{gmsh grating2D.pro -setstring test\_case LamellarGrating -setnumber flag\_polar 2} or by opening \texttt{grating2D.pro} in \texttt{Gmsh} and selecting \texttt{LamellarGrating} in the \texttt{Geometry} drop-down menu and \texttt{conical} in the \texttt{polarization case} drop-down menu.

\printbibliography

@article{demesy2007thefinite, 
author = {Guillaume Dem\'{e}sy and Fr\'{e}d\'{e}ric Zolla and Andr\'{e} Nicolet and Mireille Commandr\'{e} and Caroline Fossati}, 
journal = {Opt. Express}, 
keywords = {mypapers},
number = {26}, 
pages = {18089--18102}, 
publisher = {OSA},
title = {The finite element method as applied to the diffraction by an anisotropic grating}, 
volume = {15}, 
month = {Dec},
year = {2007},
%url = {http://www.opticsexpress.org/abstract.cfm?URI=oe-15-26-18089},
doi = {10.1364/OE.15.018089},
abstract = {The main goal of the method proposed in this paper is the numerical study of various kinds of anisotropic gratings deposited on isotropic substrates, without any constraint upon the diffractive pattern geometry or electromagnetic properties. To that end we propose a new FEM (Finite Element Method) formulation which rigorously deals with each infinite issue inherent to grating problems. As an example, 2D numerical experiments are presented in the cases of the diffraction of a plane wave by an anisotropic aragonite grating on silica substrate (for the two polarization cases and at normal or oblique incidence). We emphasize the interesting property that the diffracted field is non symmetric in a geometrically symmetric configuration.},
}

@Article{berenger94perfec-match-layer,
  author =	 {Jean-Pierre Bérenger},
  title =	 {A Perfectly Matched Layer for the Absorption of
                  Electromagnetic Waves},
  journal =	 {Journal of Computational Physics},
  year =	 {1994},
  OPTkey =	 {},
  volume =	 {114},
  OPTnumber =	 {},
  pages =	 {185--200},
  OPTmonth =	 {},
  OPTnote =	 {}
}

@book{jin2015finite,
  title={The finite element method in electromagnetics},
  author={Jin, Jian-Ming},
  year={2015},
  publisher={John Wiley \& Sons}
}

@book{stratton2007electromagnetic,
  title={Electromagnetic theory},
  author={Stratton, Julius Adams},
  volume={33},
  year={2007},
  publisher={John Wiley \& Sons}
}

@book{renversez2012foundations,
  title={Foundations of Photonic Crystal Fibres},
  author={Zolla, Frédéric and Renversez, Gilles and Nicolet, André and Kuhlmey, Boris and Guenneau, Sébastien and Felbacq, Didier and Argyros, Alexander and Leon-Saval, Sergio},
  year={2012},
  edition =  {2nd},
  publisher={Imperial College Press}}

@online{refractiveindex,
  author = {Polyanskiy, Mikhail},
  title = {},
  url = {https://refractiveindex.info/},
  urldate = {2022-11-01}
}

@book{palik1998handbook,
  title={Handbook of optical constants of solids},
  author={Palik, Edward D},
  volume={3},
  year={1998},
  publisher={Academic press}
}

@article{johnson_optical_1972,
	title = {Optical {Constants} of the {Noble} {Metals}},
	volume = {6},
	url = {http://link.aps.org/doi/10.1103/PhysRevB.6.4370},
	doi = {10.1103/PhysRevB.6.4370},
	number = {12},
	urldate = {2016-11-07TZ},
	journal = {Physical Review B},
	author = {Johnson, P. B. and Christy, R. W.},
	month = dec,
	year = {1972},
	pages = {4370--4379}
}

@article{demesy2010allpurpose, 
author = {Guillaume Dem\'{e}sy and Fr\'{e}d\'{e}ric Zolla and Andr\'{e} Nicolet and Mireille Commandr\'{e}}, 
journal = {J. Opt. Soc. Am. A}, 
keywords = {mypapers},
number = {4}, 
pages = {878--889}, 
publisher = {OSA},
title = {All-purpose finite element formulation for arbitrarily shaped crossed-gratings embedded in a multilayered stack}, 
volume = {27}, 
month = {Apr},
year = {2010},
%url = {http://josaa.osa.org/abstract.cfm?URI=josaa-27-4-878},
doi = {10.1364/JOSAA.27.000878},
abstract = {We propose a novel formulation of the finite element method adapted to the calculation of the vector field diffracted by an arbitrarily shaped crossed-grating embedded in a multilayered stack and illuminated by an arbitrarily polarized plane wave under oblique incidence. A complete energy balance (transmitted and reflected diffraction efficiencies and losses) is deduced from field maps. The accuracy of the proposed formulation has been tested using classical cases computed with independent methods. Moreover, to illustrate the independence of our method with respect to the shape of the diffractive object, we present the global energy balance resulting from the diffraction of a plane wave by a lossy thin torus crossed-grating. Finally, computation time and convergence as a function of the mesh refinement are discussed. As far as integrated energy values are concerned, the presented method shows a remarkable convergence even for coarse meshes.},
}

@article{demesy2009versatile, 
author = {Guillaume Dem\'{e}sy and Fr\'{e}d\'{e}ric Zolla and Andr\'{e} Nicolet and Mireille Commandr\'{e}}, 
journal = {Opt. Lett.}, 
keywords = {mypapers},
number = {14}, 
pages = {2216--2218}, 
publisher = {OSA},
title = {Versatile full-vectorial finite element model for crossed gratings}, 
volume = {34}, 
month = {Jul},
year = {2009},
%url = {http://ol.osa.org/abstract.cfm?URI=ol-34-14-2216},
doi = {10.1364/OL.34.002216},
abstract = {We demonstrate the accuracy of the finite-element method to calculate the diffraction efficiencies of an arbitrarily shaped crossed grating in a multilayered stack illuminated by an arbitrarily polarized plane wave under oblique incidence. The method has been validated by using classical cases found in the literature. Finally, to illustrate the independence of our method with respect to the shape of the diffractive object, we present the global energy balance resulting from the diffraction of a plane wave by a lossy thin torus crossed grating.},
}

@article{granet1999reformulation,
  title={Reformulation of the lamellar grating problem through the concept of adaptive spatial resolution},
  author={Granet, G{\'e}rard},
  journal={JOSA A},
  volume={16},
  number={10},
  pages={2510--2516},
  year={1999},
  publisher={Optica Publishing Group}
}

@article{fehrembach2002phenomenological,
  title={Phenomenological theory of filtering by resonant dielectric gratings},
  author={Fehrembach, Anne-Laure and Maystre, Daniel and Sentenac, Anne},
  journal={JOSA A},
  volume={19},
  number={6},
  pages={1136--1144},
  year={2002},
  publisher={Optica Publishing Group}
}

@article{gmsh,
    author = {C. Geuzaine and J.-F. Remacle},
    title  = "Gmsh: a three-dimensional finite element mesh generator with built-in pre- and post-processing facilities",
    journal= "International Journal for Numerical Methods in Engineering",
    volume = "79",
    number = "11",
    pages  = "1309--1331",
    year   = "2009"}

@article{getdp,
    author = {P. Dular and C. Geuzaine and F. Henrotte and W. Legros},
    title  = "A general environment for the treatment of discrete problems and its application to the finite element method",
    journal= "IEEE Transactions on Magnetics",
    volume = " 34",
    number = "5",
    pages  = "3395--3398",
    year   = "1998"}

@article{petit1980electromagnetic,
  title={Electromagnetic Theory of Gratings},
  author={Petit, Roger},
  journal={Electromagnetic Theory of Gratings. Series: Topics in Current Physics},
  volume={22},
  year={1980}
}

@book{petit_ondes_1992,
	title = {Ondes électromagnétiques en radioélectricité et en optique},
	isbn = {978-2-225-81571-3},
	language = {fr},
	publisher = {Masson},
	author = {Petit, R.},
	year = {1992}
}

@misc{onelab,
		title = {{ONELAB} website},
    url = {http://onelab.info/},
    year = {2018}
}
\end{document}